\documentclass[11pt]{article}
\usepackage{amsmath}
\usepackage{amssymb}
\usepackage{fancybox}
\usepackage{graphicx,psfrag,epsf}
\usepackage{enumerate}
\usepackage{multirow}
\usepackage{epsfig}
\usepackage{subfigure}
\usepackage{theorem}
\usepackage{natbib}
\usepackage{cite}
\usepackage{enumitem}
\usepackage{xparse}
\usepackage{makeidx}
\usepackage{chngcntr}
\usepackage{mathrsfs}

\newcommand{\blind}{0}
\newtheorem{theorem}{Theorem}[section]
\newtheorem{lemma}[theorem]{Lemma}
\newenvironment{proof}[1][Proof:]{\begin{trivlist}
\item[\hskip \labelsep {\bfseries #1}]}{\end{trivlist}}

\newcommand*{\QEDA}{\ensuremath{\blacksquare}}

\setdescription{font=\normalfont}

\DeclareDocumentEnvironment{steps}%
{O{Step}}
{\begin{enumerate}[label=#1 \arabic*,leftmargin=39pt]}%
{\end{enumerate}}

\makeatletter
\def\step{%
  \@ifnextchar[\@step{\@noitemargtrue\@step[\@itemlabel]}} 
\def\@step[#1]{\item[#1:]}
\makeatother

\begin{document}

\def\spacingset#1{\renewcommand{\baselinestretch}%
{#1}\small\normalsize} \spacingset{1}

\if0\blind
{
  \title{\bf Metropolis-Hastings within\\ Partially Collapsed Gibbs Samplers}
  \author{David A. van Dyk and Xiyun Jiao\thanks{Professor David A. van Dyk holds a Chair in Statistics in the Department of Mathematics at Imperial College London, SW7 2AZ (dvandyk@imperial.ac.uk); Xiyun Jiao is a postgraduate student in Statistics at Imperial College.}}
 \date{}
  \maketitle
} \fi

\if1\blind
{
  \bigskip
  \bigskip
  \bigskip
  \begin{center}
    {\LARGE\bf Metropolis-Hastings Algorithm within
    \\ \bigskip
    Partially Collapsed Gibbs Samplers}
  \end{center}
  \medskip
} \fi

\vspace*{-0.02in}
\begin{abstract}
The Partially Collapsed Gibbs (PCG) sampler offers a new strategy for improving the convergence of a Gibbs sampler. PCG achieves faster convergence by reducing the conditioning in some of the draws of its parent Gibbs sampler. Although this can significantly improve convergence, care must be taken to ensure that the stationary distribution is preserved. The conditional distributions sampled in a PCG sampler may be incompatible and permuting their order may upset the stationary distribution of the chain. Extra care must be taken when Metropolis-Hastings (MH) updates are used in some or all of the updates. Reducing the conditioning in an MH within Gibbs sampler can change the stationary distribution, even when the PCG sampler would work perfectly if MH were not used. In fact, a number of samplers of this sort that have been advocated in the literature do not actually have the target stationary distributions. In this article, we illustrate the challenges that may arise when using MH within a PCG sampler and develop a general strategy for using such updates while maintaining the desired stationary distribution. Theoretical arguments provide guidance when choosing between different MH within PCG sampling schemes. Finally we illustrate the MH within PCG sampler and its computational advantage using several examples from our applied work.
\end{abstract}

\noindent%

\spacingset{1.45}
\section{Introduction}
\label{sec:intro}

The popularity of the Gibbs sampler stems from its simplicity and power to effectively generate samples from a high-dimensional probability distribution. It can sometimes, however, be very slow to converge, especially when it is used to fit highly structured or complex models. The Partially  Collapsed Gibbs (PCG) sampler offers a strategy for improving the convergence characteristics of a Gibbs sampler \citep{vand:park:08,park:vand:09,vand:park:11}. A PCG sampler achieves faster convergence by reducing the conditioning in some or all of the component draws of its parent Gibbs sampler. That is, one or more of the complete conditional distributions is replaced by the corresponding complete conditional distribution of a multivariate marginal distribution of the target. For example, we might consider sampling $p({\psi}_{1}|{\psi}_{2})$ rather than $p({\psi}_{1}|{\psi}_{2},{\psi}_{3})$, where $p({\psi}_{1}|{\psi}_{2})$ is a conditional distribution of the marginal distribution, $p({\psi}_{1},{\psi}_{2})$, of the target $p({\psi}_{1},{\psi}_{2},{\psi}_{3})$. This strategy has already been proven useful in improving the convergence properties of numerous samplers \citep[e.g.,][etc.]{ber:gay:13,berr:cald:12,car:teh:14,dobi:tour:10,hans:etal:12,hu:gram:lian:12,hu:lian:13,kail:etal:10,kail:etal:11,
lin:tour:10,lin:sch:13,park:etal:08,park:vand:09,park:11,park:jeonl:lee:12,park:kraf:sanc:12,
zhao:lian:13}.

Although the PCG sampler can be very efficient, it must be implemented with care to make sure that the stationary distribution of the resulting sampler is indeed the target. Unlike the ordinary Gibbs sampler, the conditional distributions sampled in a PCG sampler may be incompatible, meaning there is no joint distribution of which they are simultaneously the conditional distributions. In this case, permuting the order of the updates can change the stationary distribution of the chain. 

As with an ordinary Gibbs sampler, we sometimes find that one or more of the conditional draws of a PCG sampler is not available in closed form and we may consider implementing such draws with the help of a Metropolis-Hastings (MH) sampler. Reducing the conditioning in one draw of an MH within Gibbs sampler, however, may alter the stationary distribution of the chain. This can happen even when the PCG sampler would work perfectly well if all of the conditional updates were available without resorting to MH updates. Examples arise even in a two-step MH within PCG sampler. \citet{wood:etal:12}, for example, points out this problem in certain samplers described in the literature for regression with functional predictors. Although they do not use the framework of PCG, these samplers are simple special cases of improper MH within PCG samplers. They first analyze the functional predictors in isolation of the regression and then use MH to update the regression parameters conditional on parameters describing the functional predictors. The first step effectively samples the functional parameters marginally and the second uses MH for sampling from the complete conditional of the regression parameters. In this article we pay special attention to this situation because it is both conceptually simple and important in practice. In Section~\ref{sec:mhrs} we propose two simple strategies that maintain the target distribution and in Section~\ref{sec:the} we compare the performance of the two strategies theoretically.

In this article, we illustrate difficulties that may arise when using MH updates within a PCG sampler and develop a general strategy for using such updates while maintaining the target stationary distribution. We begin in Section~\ref{sec:examp} with two motivating examples that are chosen to review the subtleties of the PCG sampler, illustrate the complications that arise when MH is introduced into PCG, and set the stage for the methodological and theoretical contributions of this article. Section~\ref{sec:examp} ends by reviewing the method of \citet{vand:park:08} for establishing the stationary distribution of a PCG sampler. The MH within PCG sampler is introduced in Section~\ref{sec:mhpcg} along with methods for ensuring that its stationary distribution is the target distribution and several strategies for implementing the sampler while maintaining this target. Theoretical arguments are presented in Section~\ref{sec:the} that aim to guide the choice between different implementations of the MH within PCG sampler. The proposed methods and theoretical results are illustrated in Section~\ref{sec:exa} in the context of several examples, including factor analysis and two examples from high-energy astrophysics. The factor analysis example contrasts the step-ordering constraints of MH within PCG and of the related ECME algorithm \citep{liu:rubi:94}. Final discussion appears in Section~\ref{sec:disc}. 

\section{Background and Motivating Examples}
\label{sec:examp}

\subsection{Notation}
\label{sec:note}

We aim to sample from the target distribution, $p(\psi)$, by constructing a Markov chain \{${\psi}^{(t)}, t=1,2,\dots$\} with the stationary distribution $\pi(\psi)$, where $\psi$ is a multivariate random variable. That is, we aim to construct a Markov chain such that $\pi(\psi)=p(\psi)$. We refer to a sampler as {\it proper} if it has a stationary distribution and that distribution coincides with the target, i.e., $\pi(\psi)=p(\psi)$; otherwise we call the sampler {\it improper}. Typically $p(\psi)$ is the posterior distribution in a Bayesian analysis, but this is not necessary. In data-driven examples, we use standard Bayesian notation. 

To facilitate discussion of the relevant samplers, we divide $\psi$ into $J$ possibly multivariate {non-overlapping} subcomponents, i.e., $\psi=({\psi}_{1},\dots,{\psi}_{J})$, and define $\mathscr{J}=\{1,2,\dots,J\}$. {The methods that we consider are Gibbs-type samplers that rely on the conditional distributions of either $p(\psi)$ or its multivariate marginal distributions. When conditional distributions cannot be sampled directly, we may use MH. For example, suppose we wish to sample the conditional distribution $p(\psi_{j_1}|\psi_{j_2})$ of the marginal distribution $p(\psi_{j_1},\psi_{j_2})$, but cannot do so directly. In this case, we specify a jumping rule (i.e., a proposal distribution), denoted by $\mathcal{J}_{j_{1}|j_{2}}({\psi}_{j_{1}}|{\psi}_{j_{1}}^\prime,{\psi}_{j_{2}}^\prime,{\psi}_{j_{3}}^\prime)$, where the subscript specifies the target conditional distribution and we use primes to indicate the current value of the subcomponents of $\psi$; notice that the jumping rule may depend on subcomponents other than $\psi_{j_1}^\prime$ and $\psi_{j_2}^\prime$, namely, $\psi_{j_3}^\prime$. In the MH update, we sample $\psi_{j_1}^{\rm prop}\sim\mathcal{J}_{j_{1}|j_{2}}({\psi}_{j_{1}}|{\psi}_{j_{1}}^\prime,{\psi}_{j_{2}}^\prime,{\psi}_{j_{3}}^\prime)$ and set $\psi_{j_1}=\psi_{j_1}^{\rm prop}$ with probability $r=\mbox{min}\left\{1, \ \displaystyle{\frac{p(\psi_{j_1}^{\rm prop} | \psi_{j_2}^\prime)\mathcal{J}_{j_{1}|j_{2}}({\psi}_{j_1}^\prime|{\psi}_{j_{1}}^{\rm prop},{\psi}_{j_{2}}^\prime,{\psi}_{j_{3}}^\prime)}{p(\psi_{j_1}^\prime|\psi_{j_2}^\prime)\mathcal{J}_{j_{1}|j_{2}}({\psi}_{j_{1}}^{\rm prop}|{\psi}_{j_{1}}^\prime,{\psi}_{j_{2}}^\prime,{\psi}_{j_{3}}^\prime)}}\right\}$; otherwise the current value is retained, i.e., $\psi_{j_1}=\psi_{j_1}^\prime$. This MH transition kernel, denoted by $\mathcal{M}_{j_{1}|j_{2}}({\psi}_{j_{1}}|{\psi}_{j_{1}}^\prime,{\psi}_{j_{2}}^\prime,{\psi}_{j_{3}}^\prime)$, has stationary distribution $p({\psi}_{j_{1}}|{\psi}_{j_{2}})$. We can also express the iterates explicitly. For instance, ${\psi}_{2}^{(t+1)}\sim\mathcal{M}_{2|1,3}({\psi}_{2}|{\psi}_{1}^{(t+1)},{\psi}_{2}^{(t)},{\psi}_{3}^{(t)})$ is a typical expression for sampling from an MH transition kernel with stationary distribution $p({\psi}_{2}|{\psi}_{1}^{(t+1)},{\psi}_{3}^{(t)})$. Notice that this transition kernel depends on ${\psi}_{2}^{(t)}$ because the acceptance probability involves ${\psi}_{2}^{(t)}$ and because ${\psi}_{2}^{(t+1)}$ is set to ${\psi}_{2}^{(t)}$ if the proposal is rejected.} Here we introduce two examples that illustrate the advantages and potential pitfalls that may arise when using PCG samplers when MH is required for some of their updates.

\subsection{Spectral analysis in X-ray astronomy}
\label{sec:saxa}

We begin with an example from our applied work in X-ray astronomy that involves a spectral analysis model that can be fitted with the Data Augmentation algorithm and Gibbs-type samplers \citep{vand:conn:kash:siem:01,vand:meng:10}. We use variants of this example as a running illustration of the methods we propose. The X-ray detectors used in astronomy are typically on board space-based observatories and record the number of photons detected in each of a large number of energy bins. Spectral analysis aims to estimate the distribution of the photon energies. We use Poisson models for the recorded photon counts, where the expected count is parameterized as a function of the energy, $E_{i}$ of bin $i$. A simple example is
\begin{eqnarray}
 X_{i}\stackrel{\mbox{\tiny{ind}}}{\sim}{\rm Poisson}\bigg\{{\Lambda}_{i}=\alpha({E_{i}}^{-\beta}+{\gamma}I\{i=\mu\})e^{-\phi/E_{i}}\bigg\},\mbox{ for } i=1,\dots,n,
\label{eq:sesa}
\end{eqnarray}
where $X_{i}$ is the count in bin $i$; $\alpha$, $\beta$, $\gamma$, $\mu$ and $\phi$ are model  parameters; $I\{\cdot\}$ is the indicator function; and $n$ is the number of energy bins. The $\alpha{E_{i}}^{-\beta}$ term in~(\ref{eq:sesa}) is a {\it continuum}---a smooth term that extends over a wide range of energies. The $\alpha\gamma I\{i=\mu\}$ term is an {\it emission line}---a sharp narrow term that describes a distinct aberration from the continuum. The emission line in~(\ref{eq:sesa}) is very narrow in that it is contained entirely in one energy bin. The parameters of the continuum and emission line describe the composition, temperature, and general physical environment of the source. The factor $e^{-\phi/E_{i}}$ in~(\ref{eq:sesa}) accounts for absorption---lower energy photons are more likely to be absorbed by inter-stellar material and not be recorded by the detector. A typical spectral model might contain multiple summed continua and emission lines. We use a simple example here to focus attention on computational issues. Since $\alpha$, $\beta$, $\gamma$ and $\phi$ {are often} blocked in the samplers we discuss, we refer to them jointly as $\theta=(\alpha,\beta,\gamma,\phi)$. We assume that $\theta$ and $\mu$ are {\it a priori} independent and that $\mu$ is {\it a priori} uniform on $\{1,\dots,n\}$.

In practice, we do not observe $X=(X_{1},\dots,X_{n})$ directly because photon counts are subject to stochastic censoring, misclassification, and background contamination. First,  because the sensitivity of the detector varies with energy, the probability that a photon is detected depends on its energy. Combining this with background contamination,
\begin{eqnarray}
\tilde X_i \mid X_i \stackrel{\mbox{\tiny{ind}}}{\sim}{\rm Binomial}\big\{X_i,A_i\big\}+{\rm Poisson}(\xi_i), \ \mbox{ for } \  i=1,\dots,n,
\label{eq:nusesa}
\end{eqnarray}
where ${\tilde X}=(\tilde X_{1},\dots,\tilde X_{n})$  are the photon counts, including background, that are not absorbed, $A=(A_{1},\dots,A_{n})$ is the {\it effective area} of the detector which describes its sensitivity, and $\xi=(\xi_{1},\dots,\xi_{n})$ is the expected background count. Second, misclassification occurs because a photon with energy $E_i$ has probability $P_{ij}$ of being recorded in bin $j$. Combining these effects, the conditional distribution of the observed photon counts $Y=(Y_{1},\dots,Y_{n})$ given ${\tilde X}$ is 
\begin{eqnarray}
Y \mid {\tilde X} 
\stackrel{\mbox{\tiny{ind}}}{\sim} 
\sum_{i=1}^n {\rm Multinomial}\bigg\{ \tilde X_i, \ (P_{i1},\dots,P_{in})\bigg\},
\label{eq:nusa}
\end{eqnarray}
and marginally,
\begin{eqnarray}
Y_{j}\stackrel{\mbox{\tiny{ind}}}{\sim}{\rm Poisson}\bigg\{ \displaystyle \sum\limits_{i=1}^{n}P_{ij}(A_i {\Lambda}_{i}+{\xi}_{i})\bigg\}, \mbox{ for }j=1,\dots,n,
\label{eq:sa}
\end{eqnarray}
where $\Lambda_{i}$ is given by~(\ref{eq:sesa}). 
While $A$ and $P=\{P_{ij}\}$ are typically assumed known from instrumental calibration (see~\citeauthor{lee:etal:11},~\citeyear{lee:etal:11}, for an exception), $\xi$ is often specified in terms of a number of unknown parameters.

The model in~(\ref{eq:sesa}) is a finite mixture model and can be fitted via the standard data augmentation scheme that sets $X_{i}=X_{iC}+X_{iL}$, where $X_{iC}\stackrel{\mbox{\tiny{ind}}}{\sim}{\rm Poisson}\left(\alpha{E_{i}}^{-\beta}e^{-\phi/E_{i}}\right)$ and $X_{iL}\stackrel{\mbox{\tiny{ind}}}{\sim}{\rm Poisson}\left(\alpha{\gamma}I\{i=\mu\}e^{-\phi/E_{i}}\right)$, are the photon counts in bin $i$ generated from the continuum and emission line, respectively. We consider samplers that target $p(X, X_L, \theta, \mu |Y)$ rather than $p(\theta, \mu |Y)$ both because the ideal data, $X$, is of scientific interest and because its introduction simplifies the complete conditional distributions, especially in more complex models with multiple summed continua and spectral lines. Assuming $\xi$ is known, this leads to a Gibbs sampler for (\ref{eq:sesa})--(\ref{eq:sa}):

\begin{steps}
\itemsep=0in
\step $(X^{(t+1)},X_L^{(t+1)})\sim p(X,X_L|Y,\theta^{(t)},\mu^{(t)})$,\hfill(Sampler 1)
\step $\theta^{(t+1)}\sim p(\theta|Y,X^{(t+1)},X_L^{(t+1)},\mu^{(t)})$,
\step $\mu^{(t+1)}\sim p(\mu|Y,X^{(t+1)},X_L^{(t+1)},\theta^{(t+1)})$,
\end{steps}

\noindent  where $X_L=(X_{1L},\dots,X_{nL})$. We separate $\mu$ and $\theta$ into two steps to facilitate derivation of the partially collapsed versions of this sampler. Because $X_L$ completely specifies the line location, $\mu$, ${\rm Var}_{\pi}(\mu|X_L)=0$, Sampler~1 is not irreducible, and $\mu^{(t)}=\mu^{(0)}$ for all $t$, for any choice of $\mu^{(0)}$. This problem can be solved by updating $\mu$ without conditioning on $X_L$. In particular, we can replace Step~3 of Sampler~1 with $(X_L^{(t+1)},\mu^{(t+1)})\sim p(X_L,\mu|Y,X^{(t+1)},\theta^{(t+1)})$ and permute the steps to

\begin{steps}
\itemsep=0in
\step $(X_L^{*},\mu^{(t+1)})\sim p(X_L,\mu|Y,X^{(t)},\theta^{(t)})$,\hfill(Sampler 2)
\step $(X^{(t+1)},X_L^{(t+1)})\sim p(X,X_L|Y,\theta^{(t)},\mu^{(t+1)})$,
\step $\theta^{(t+1)}\sim p(\theta|Y,X^{(t+1)},X_L^{(t+1)},\mu^{(t+1)})$.
\end{steps}

\noindent The sampled $X_L$ in Step~1 is denoted by $X_L^{*}$ because it is not an output of the Markov transition kernel; $X_L$ is updated again in Step~2. In fact $X_L^{*}$ is a redundant quantity in that it is not used at all subsequent to Step~1 and replacing Step~1 with $\mu^{(t+1)} \sim p(\mu|Y,X^{(t)},\theta^{(t)})$ does not alter the Markov transition kernel of Sampler~2. The resulting sampler, that is,
\begin{steps}
\itemsep=0in
\step $\mu^{(t+1)}\sim p(\mu|Y,X^{(t)},\theta^{(t)})$,\hfill(Sampler 3)
\step $(X^{(t+1)},X_L^{(t+1)})\sim p(X,X_L|Y,\theta^{(t)},\mu^{(t+1)})$,
\step $\theta^{(t+1)}\sim p(\theta|Y,X^{(t+1)},X_L^{(t+1)},\mu^{(t+1)})$,
\end{steps}
\noindent is an example of a PCG sampler composed of incompatible conditional distributions. A variant of this sampler was discussed in~\citet{park:vand:09}.

By its construction, the stationary distribution of Sampler~3 is $p(X,X_L,\theta,\mu|Y)$, see Section~\ref{sec:con}. Unlike an ordinary Gibbs sampler, however, permuting its steps may alter its stationary distribution. Suppose, for example, we obtain $(X^{(t)},X_L^{(t)},\theta^{(t)},\mu^{(t)})$ from $p(X,X_L,\theta,\mu|Y)$ and update $\mu$ according to Step~1 of Sampler~3. The joint distribution of $(X^{(t)},X_L^{(t)},\theta^{(t)},\mu^{(t+1)})$ would be
\begin{eqnarray}
\int p(\mu^{(t+1)}|Y,X^{(t)},\theta^{(t)})p(X^{(t)},X_L^{(t)},\theta^{(t)},\mu^{(t)}|Y)d\mu^{(t)}=p(X^{(t)},\theta^{(t)},\mu^{(t+1)}|Y)p(X_L^{(t)}|Y,X^{(t)},\theta^{(t)}).
\label{eq:inter}
\end{eqnarray}
It is the conditional independence of $X_L^{(t)}$ and $\mu^{(t+1)}$ in~(\ref{eq:inter}) that makes Sampler~3 so much faster than Sampler~1; recall ${\rm Var}_{\pi}(\mu|X_L)=0$. Because the joint distribution of $\theta^{(t)}$ and $\mu^{(t+1)}$ in~(\ref{eq:inter}) is their posterior distribution and Step~2 conditions only on $\theta^{(t)}$ and $\mu^{(t+1)}$, the joint distribution of the unknowns after Step~2, that is, of $(X^{(t+1)},X_L^{(t+1)},\theta^{(t)},\mu^{(t+1)})$, is again the target posterior. Thus a cyclic permutation of the steps in Sampler~3 that ends either with Step~2 or Step~3 results in a proper sampler, but ending with Step~1 does not. With non-cyclic permutations, the stationary distribution is unknown. 

\subsection{A common error in the simplest PCG sampler}
\label{sec:cespcg}

The potential pitfalls of introducing MH updates into a PCG sampler can be illustrated using the simplest possible PCG sampler. To see this, we start with a two-step Gibbs sampler with target distribution $p({\psi}_{1},{\psi}_{2})$, where the second step relies on an MH update:
\begin{steps}
\itemsep=0in
\step ${\psi}_{1}^{(t+1)} \sim p({\psi}_{1}|{\psi}_{2}^{(t)})$,\hfill(Sampler 4)
\step ${\psi}_{2}^{(t+1)}\sim{{\mathcal{M}}_{2|1}({\psi}_{2}|{\psi}_{1}^{(t+1)},\psi_2^{(t)})}$.
\end{steps}
\noindent While this sampler is proper, replacing Step 1 with ${\psi}_{1}^{(t+1)}\sim p({\psi}_{1})$ results in an improper sampler:
\begin{steps}
\itemsep=0in
\step ${\psi}_{1}^{(t+1)}\sim p({\psi}_{1})$,\hfill(Sampler 5)
\step ${\psi}_{2}^{(t+1)}\sim{{\mathcal{M}}_{2|1}({\psi}_{2}|{\psi}_{1}^{(t+1)},\psi_2^{(t)})}$.
\end{steps}

\begin{figure}[t]
\spacingset{1}
\begin{center}
  \includegraphics[width=6in]{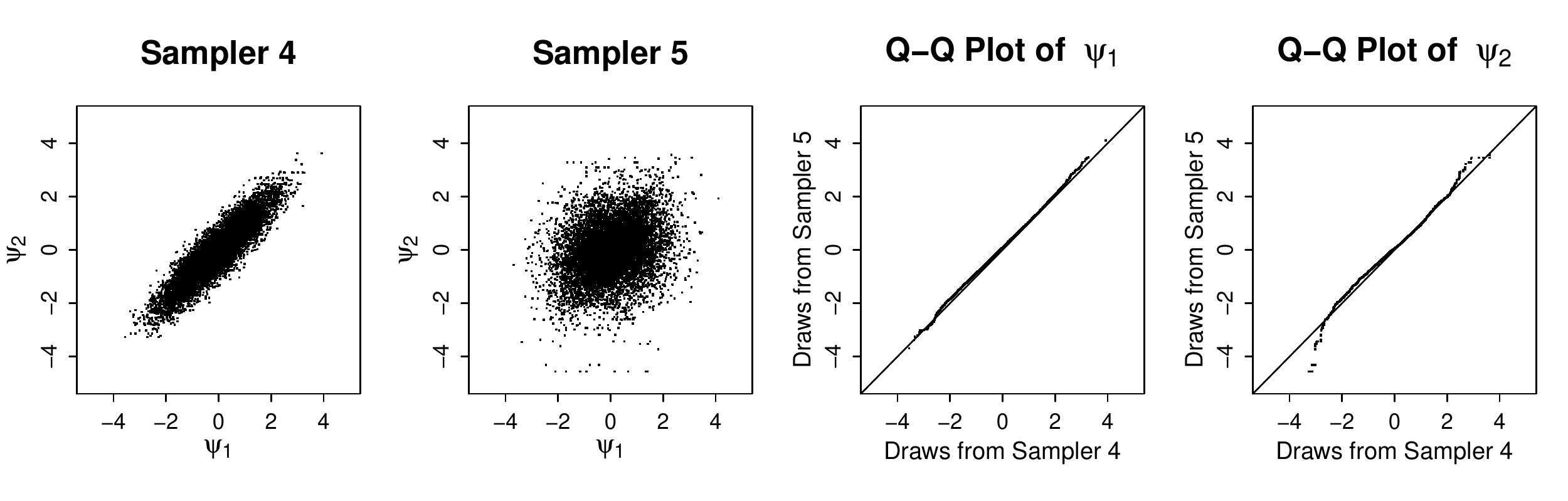}
    \caption{Proper and improper samplers, for the bivariate normal target distribution. The first two panels give scatter plots of $\psi_1$ and $\psi_2$ for 10,000 draws from Samplers~4 and~5, respectively. The marginal distributions of the two samplers are compared in the two quantile-quantile plots. The improper Sampler 5 severely underestimates the correlation between $\psi_1$ and $\psi_2$, and slightly overestimates the variance of $\psi_{2}$.}
    \label{fig:pitfall}
\end{center}
\end{figure}  

\noindent The problem with Sampler~5 can be illustrated using a simulation study. Figure~\ref{fig:pitfall} compares 10,000 draws generated by Samplers~4 and~5 with $p({\psi}_{1},{\psi}_{2})$ given by 
\begin{eqnarray}
\left(\begin{array}{c}
{\psi}_{1}\\
{\psi}_{2}
\end{array}
\right) \sim{\rm N}_{2}\left[\left(\begin{array}{c}
0\\
0
\end{array}
\right),\left(\begin{array}{cc}
1&0.9\\
0.9&1
\end{array}
\right)\right].
\label{eq:normal}
\end{eqnarray}
The MH jumping rule in Step~2 of both samplers is a Gaussian distribution centered at the previous draw with variance equal to 3. Sampler~5 underestimates the correlation of the target distribution and overestimates the marginal variance of $\psi_2$. {Of course, if we repeat Step 2 a sufficient number of times within each iteration of Sampler 5, it would deliver a draw (nearly) from its target, $p(\psi_2|\psi_1)$, and Sampler 5 would deliver (nearly) independent draws from $p(\psi_1,\psi_2)$. We discuss this strategy for constructing an approximately proper sampler in Section~\ref{sec:mhrs}. Similarly, iterating Step~2 of Sampler~4 would (nearly) lead to a standard two-step Gibbs sampler.}

The key to understanding the failure of Sampler~5 (without iterating Step 2) lies in the MH jumping rule used in Step~2 of both samplers. {The kernel ${\mathcal{M}}_{2|1}$ depends on ${\psi}_{2}^{(t)}$ through its acceptance probability and its output if its proposal is rejected, thus ${\mathcal{M}}_{2|1}$ must be written as ${\mathcal{M}}_{2|1}({\psi}_{2}|{\psi}_{1}^{(t+1)},{\psi}_{2}^{(t)})$. Although ${\mathcal{M}}_{2|1}$ delivers a draw from $p({\psi}_{2}|{\psi}_{1}^{(t+1)})$ if given a sample $({\psi}_{1}^{(t+1)},{\psi}_{2}^{(t)})$ from the target distribution, in Sampler~5, ${\psi}_{1}^{(t+1)}$ and ${\psi}_{2}^{(t)}$ are independent and ${\mathcal{M}}_{2|1}$ does not deliver a draw from $p({\psi}_{2}|{\psi}_{1}^{(t+1)})$.}

Unfortunately, there are several examples of samplers in the literature that have the same structure as the improper Sampler 5, for instance, \citet{liu:etal:09}, \citet{lunn:etal:09}, \citet{mccan:etal:10}, and even in the popular WinBUGS package (Spiegelhalter, Thomas, Best and Lunn 2003), see Section~\ref{sec:smhpcg}. These samplers do not generally exhibit the desired stationary distributions.

\subsection{Convergence of the Partially Collapsed Gibbs sampler}
\label{sec:con}

\begin{figure}[t]%
\centering
\subfiguretopcaptrue
\subfigure[Parent Gibbs Sampler]{%
\framebox[0.995\width]{
\begin{minipage}[c][1.6cm][c]{0.23\textwidth}
\begin{center}
\raggedright
\footnotesize{
\hfill$\psi_1\sim p(\psi_1|\psi_2^\prime,\psi_3^\prime,\psi_4^\prime)$\par
\hfill$\psi_2\sim p(\psi_2|\psi_1,\psi_3^\prime,\psi_4^\prime)$\par
$(\psi_3,\psi_4)\sim p(\psi_3,\psi_4|\psi_1,\psi_2)$
}
\end{center}
\end{minipage}
}}%
\hspace{0pt}%
\subfigure[Reduce Conditioning]{%
\framebox[0.995\width]{
\begin{minipage}[c][1.6cm][c]{0.24\textwidth}
\begin{center}
\raggedright
\footnotesize{
\hfill $(\psi_1,\psi_3^\star)\sim p(\psi_1,\psi_3|\psi_2^\prime,\psi_4^\prime)$\par
\hfill $\psi_2\sim p(\psi_2|\psi_1,\psi_3^\star,\psi_4^\prime)$\par
\hfill $(\psi_3,\psi_4)\sim p(\psi_3,\psi_4|\psi_1,\psi_2)$
}
\end{center}%
\end{minipage}
}}
\hspace{-7pt}
\subfigure[Permute]{%
\framebox[0.995\width]{
\begin{minipage}[c][1.6cm][c]{0.24\textwidth}
\begin{center}
\raggedright
\footnotesize{
\hfill $\psi_2\sim p(\psi_2|\psi_1^\prime,\psi_3^\prime,\psi_4^\prime)$\par
\hfill $(\psi_1,\psi_3^\star)\sim p(\psi_1,\psi_3|\psi_2,\psi_4^\prime)$\par
\hfill $(\psi_3,\psi_4)\sim p(\psi_3,\psi_4|\psi_1,\psi_2)$
}
\end{center}
\end{minipage}
}}%
\hspace{0pt}%
\subfigure[Trim]{%
\framebox[0.995\width]{
\begin{minipage}[c][1.6cm][c]{0.23\textwidth}
\begin{center}
\raggedright
\footnotesize{
\hfill$\psi_2\sim p(\psi_2|\psi_1^\prime,\psi_3^\prime,\psi_4^\prime)$\par
\hskip 22pt $\psi_1\sim p(\psi_1|\psi_2,\psi_4^\prime)$\par
$(\psi_3,\psi_4)\sim p(\psi_3,\psi_4|\psi_1,\psi_2)$
}
\end{center}
\end{minipage}
}}
\caption{A three-phase framework for deriving a proper PCG sampler.  The parent Gibbs sampler appears in (a). The sampler in (b) reduces the conditioning in Step~1 by updating  $\psi_3$ rather than conditioning on it. The steps of this sampler are permuted in (c) to allow the redundant draw of $\psi_3^{\star}$---in Step~2 of (c)---to be trimmed in the PCG sampler in (d).}%
\label{fig:pcg}%
\end{figure}

A three-phase framework for deriving proper PCG samplers is given in~\citet{vand:park:08}. { Consider the Gibbs sampler in Figure~\ref{fig:pcg}(a) that updates the components of~$\psi=({\psi}_{1},\psi_2,\psi_3,{\psi}_{4})$~in three steps. In the first phase of the framework, one or more steps of the parent Gibbs sampler are replaced by steps that update rather than condition upon some components of~$\psi$. This is illustrated in Figure~\ref{fig:pcg}(b), where the update $\psi_1\sim p(\psi_1|\psi_2^\prime,\psi_3^\prime,\psi_4^\prime)$ in Step~1 is replaced with~$(\psi_1,\psi_3^\star)\sim p(\psi_1,\psi_3|\psi_2^\prime,\psi_4^\prime)$. Notice that in the modified step,~$\psi_3$~is sampled rather than conditioned upon.} This {\it conditioning reduction} phase is key to the improved convergence properties of the PCG sampler. By conditioning on less, we expect to increase the variance of the updating distribution, at least on average. This is evident in Section~\ref{sec:saxa} where the complete conditional for $\mu$ in Sampler~1 has zero variance, but its update with reduced conditioning in Sampler~2 readily allows $\mu$ to move across its parameter space. More formally,~\citet{vand:park:08}~showed that sampling more unknowns in any set of steps of a Gibbs sampler can only reduce the so-called cyclic-permutation bound on the spectral radius of the sampler. The resulting substantial improvement in the rate of convergence is illustrated in the examples given in~\citet{ber:gay:13},~\citet{berr:cald:12},~\citet{car:teh:14},~\citet{dobi:tour:10},~\citet{hu:gram:lian:12},~\citet{hu:lian:13},~\citet{kail:etal:10,kail:etal:11},
~\citet{lin:tour:10},~\citet{lin:sch:13},~\citet{park:etal:08},~\citet{park:vand:09},~\citet{park:jeonl:lee:12},~\citet{park:kraf:sanc:12},
and~\citet{zhao:lian:13}, etc. ({\it Conditioning reduction} was called {\it marginalization} by \citet{vand:park:08}.)

The conditioning reduction phase results in one or more components of $\psi$ being updated in multiple steps; $\psi_3$ is updated in Steps~1 and 3 in Figure~\ref{fig:pcg}(b).
If the same component is updated in two consecutive steps, the Markov transition kernel does not depend on the first update. We call quantities that are updated in a sampler, but do not affect its transition kernel {\it redundant quantities}---they must be updated subsequently or they would be part of the output of the iteration. The second phase of the framework is to {\it permute} the steps of the sampler with reduced conditioning to make as many of the updates redundant as possible.  {For example, we permuted the steps in Figure~\ref{fig:pcg}(b) so that $\psi_3$ is updated in Steps~2 and~3 of Figure~\ref{fig:pcg}(c) and $\psi_3^\star$ is redundant.

{In the third phase, redundant quantities are removed or {\it trimmed} from the updating scheme. For example, Step~2 in Figure~\ref{fig:pcg}(d) does not update $\psi_3$. 
By construction, this does not affect the overall transition kernel. The resulting step samples from a conditional distribution of a marginal distribution of $p(\psi)$. For example, Step~2
in Figure~\ref{fig:pcg}(d)  simulates from a conditional distribution of $p(\psi_1, \psi_2,\psi_4)$ rather than of $p(\psi_1, \psi_2,\psi_3,\psi_4)$. We refer to steps that sample or target such distributions as {\it reduced steps} and to steps that sample or target a complete conditional as {\it full steps}. 

In some cases, the result of the three-phase framework is simply a blocked or collapsed~\citep{liu:wong:kong:94} version of the parent Gibbs sampler. In other cases, however, the resulting PCG sampler is composed of samples from a set of incompatible conditional distributions (e.g., Sampler~3). Since all three phases preserve the stationary distribution of the parent sampler, we know that the resulting PCG sampler is proper. Because reducing the conditioning can significantly improve the rate of convergence of the sampler, while permutation typically has a minor effect, and trimming has no effect on the rate of convergence, we generally expect the PCG sampler to exhibit better and often much better convergence properties than its parent Gibbs sampler.


\section{Using MH Algorithm within the PCG Sampler}
\label{sec:mhpcg}

\subsection{Identifying the stationary distributions}
\label{sec:idsd}

We now consider the use of MH updates for some of the steps of a PCG sampler. As the example in Section~\ref{sec:cespcg} illustrates, introducing MH into a well behaved PCG sampler can destroy the sampler's stationary distribution. Thus, care must be taken to guarantee that an MH within PCG sampler is proper. Here we describe the basic complication that arises when MH is introduced into a PCG sampler and give advice as to how to ensure that the sampler is proper. 

When deriving a PCG sampler (without MH), the conditioning reduction phase means some components of $\psi$ are updated in multiple steps. If the same component is updated in consecutive steps, the Markov transition kernel does not depend on the first update. The first update is therefore redundant and can be omitted without affecting the stationary distribution of the chain. 

This situation is more complicated when some of the steps of the PCG sampler require MH updates. Suppose, for example, we wish to sample from $p(\psi)$ with $\psi=({\psi}_{1},{\psi}_{2},{\psi}_{3})$ using a proper PCG sampler in which $\psi_1$ and $\psi_2$ are jointly updated in Step~$K$ via a draw from the conditional distribution $p({\psi}_{1},{\psi}_{2}|{\psi}_{3})$. Suppose also that $\psi_2$ is to be updated according to its full conditional distribution, $p({\psi}_{2}|{\psi}_{1},{\psi}_{3})$ in Step~$K+1$, but this cannot be done directly and we wish to use an MH update. The remaining unknowns, $\psi_3$, are updated in other steps of the sampler, which perhaps involve dividing $\psi_3$ into multiple subcomponents. That is, Steps $K$ and~$K+1$ of the sampler are
\begin{description}
\itemsep=0in
\item[Step $K$:] $(\psi_1^{(t+1)},\psi_2^{*}) \sim p(\psi_1,\psi_2|\psi_3^\prime)$,\hfill(Sampler Fragment 1)
\item[Step $K+1$:] $\psi_2^{(t+1)} \sim {\mathcal M}_{2|1,3}(\psi_2|\psi_1^{(t+1)},\psi_2^{*},\psi_3^\prime)$.
\end{description}
\noindent If we were able to draw $\psi_2$ directly from its complete conditional distribution in Step~$K+1$, $\psi_2^{*}$ would be redundant and we could remove it from the sampler by replacing the update in Step~$K$ with the reduced step $\psi_1^{(t+1)} \sim p(\psi_1|\psi_3^\prime)$. The MH update in Step~$K+1$, however, depends on $\psi_2^{*}$ and replacing it with $\psi_2^{(t)}$ may change the chain's stationary distribution in an unpredictable way. In short, the MH update used in Step~$K+1$ means that we cannot reduce Step~$K$. 
Generally speaking,  an MH update in a step that follows a reduced step is problematic because reduced steps result in independences that do not exist in the target. (A reduced step that follows an MH step, however, is not inherently problematic.)
{More precisely, the kernel, $\mathcal{M}_{j_{1}|j_{2}}({\psi}_{j_{1}}|{\psi}_{j_{1}}^\prime,{\psi}_{j_{2}}^\prime,{\psi}_{j_{3}}^\prime)$, can only be used if no component of $({\psi}_{j_{1}},{\psi}_{j_{2}},{\psi}_{j_{3}})$ is trimmed in the previous step.}

\begin{figure}[t]%
\centering
\subfiguretopcaptrue
\subfigure[Parent MH within Gibbs Sampler]{%
\fbox{
\begin{minipage}[c][3.7cm][c]{0.46\linewidth}
\begin{center}
\begin{steps}
\itemsep=-0.1in
\step ${p(X_{L}|X,\alpha^\prime,\beta^\prime,\gamma^\prime,\mu^\prime,\phi^\prime)}$
\step $p(\alpha|X,X_{L},\beta^\prime,\gamma^\prime,\mu^\prime,\phi^\prime)$
\step ${\mathcal{M}_{\beta|X,X_{L},\alpha,\gamma,\mu,\phi}(\beta|X_{L},\alpha,\beta^\prime,\gamma^\prime,\mu^\prime,\phi^\prime)}$
\step ${p(\gamma|X,X_{L},\alpha,\beta,\mu^\prime,\phi^\prime)}$
\step ${\mathcal{M}_{\mu|X,X_{L},\alpha,\beta,\gamma,\phi}(\mu|X_{L},\alpha,\beta,\gamma,\mu^\prime,\phi^\prime)}$
\step ${\mathcal{M}_{\phi|X,X_{L},\alpha,\beta,\gamma,\mu}(\phi|X_{L},\alpha,\beta,\gamma,\mu,\phi^\prime)}$
\end{steps}
\end{center}
\end{minipage}
}}%
\hspace{5pt}%
\subfigure[Reduce Conditioning]{%
\fbox{
\begin{minipage}[c][3.7cm][c]{0.46\linewidth}
 \begin{center}
\begin{steps}
\itemsep=-0.1in
\step ${p(X_{L}^\star|X,\alpha^\prime,\beta^\prime,\gamma^\prime,\mu^\prime,\phi^\prime)}$
\step $p(\alpha^\star,X_{L}^\star|X,\beta^\prime,\gamma^\prime,\mu^\prime,\phi^\prime)$
\step ${\mathcal{M}^\star_{\beta,X_{L},\alpha|X,\gamma,\mu,\phi}(\beta,X_{L}^\star,\alpha^\star|\beta^\prime,\gamma^\prime,\mu^\prime,\phi^\prime)}$
\step ${p(\gamma|X,X_{L}^\star,\alpha^\star,\beta,\mu^\prime,\phi^\prime)}$
\step ${\mathcal{M}^\star_{\mu,X_{L},\alpha|X,\beta,\gamma,\phi}(\mu,X_{L}^\star,\alpha^\star|\beta,\gamma,\mu^\prime,\phi^\prime)}$
\step ${\mathcal{M}^\star_{\phi,X_{L},\alpha|X,\beta,\gamma,\mu}(\phi,X_{L},\alpha|\beta,\gamma,\mu,\phi^\prime)}$
\end{steps}
\end{center}
\end{minipage}
}}\\
\subfigure[Permute]{%
\fbox{
\begin{minipage}[c][3.7cm][c]{0.46\linewidth}
 \begin{center}
\begin{steps}
\itemsep=-0.1in
\step ${\mathcal{M}^\star_{\mu,X_{L},\alpha|X,\beta,\gamma,\phi}(\mu,X_{L}^\star,\alpha^\star|\beta^\prime,\gamma^\prime,\mu^\prime,\phi^\prime)}$
\step ${\mathcal{M}^\star_{\phi,X_{L},\alpha|X,\beta,\gamma,\mu}(\phi,X_{L}^\star,\alpha^\star|\beta^\prime,\gamma^\prime,\mu,\phi^\prime)}$
\step ${\mathcal{M}^\star_{\beta,X_{L},\alpha|X,\gamma,\mu,\phi}(\beta,X_{L}^\star,\alpha^\star|\beta^\prime,\gamma^\prime,\mu,\phi)}$
\step $p(\alpha,X_{L}^\star|X,\beta,\gamma^\prime,\mu,\phi)$
\step ${p(X_{L}|X,\alpha,\beta,\gamma^\prime,\mu,\phi)}$
\step ${p(\gamma|X,X_{L},\alpha,\beta,\mu,\phi)}$
\end{steps}
\end{center}
\end{minipage}
}}
\hspace{2pt}%
\subfigure[Trim]{%
\fbox{
\begin{minipage}[c][3.7cm][c]{0.46\linewidth}
 \begin{center}
\begin{steps}
\itemsep=-0.1in
\step ${\mathcal{M}_{\mu|X,\beta,\gamma,\phi}(\mu|\beta^\prime,\gamma^\prime,\mu^\prime,\phi^\prime)}$
\step ${\mathcal{M}_{\phi|X,\beta,\gamma,\mu}(\phi|\beta^\prime,\gamma^\prime,\mu,\phi^\prime)}$
\step ${\mathcal{M}_{\beta|X,\gamma,\mu,\phi}(\beta|\beta^\prime,\gamma^\prime,\mu,\phi)}$
\step $p(\alpha|X,\beta,\gamma^\prime,\mu,\phi)$
\step ${p(X_{L}|X,\alpha,\beta,\gamma^\prime,\mu,\phi)}$
\step ${p(\gamma|X,X_{L},\alpha,\beta,\mu,\phi)}$
\end{steps}
\end{center}
\end{minipage}
}}
\caption{Three-phase framework used to derive Sampler~6 from its parent MH within Gibbs sampler. The parent sampler appears in (a) with Steps 3, 5 and 6 requiring MH updates. The conditioning in steps 2, 3, 5, and 6 is reduced  in (b). The steps are permuted in (c) to allow redundant draws of $X_L^{\star}$ and $\alpha^{\star}$ to be trimmed in Steps 1--4. The resulting proper MH within PCG sampler, i.e., Sampler 6, appears in (d).}%
\label{fig:spectralmodel61}%
\end{figure}

Luckily, the stationary distribution of an MH within PCG sampler can be verified using the same methods that are used for an ordinary PCG sampler. In particular, the three-phase framework of~\citet{vand:park:08} can be directly applied. The first two phases, conditioning reduction and permutation, apply equally well to MH within Gibbs samplers. Neither updating additional components of $\psi$ in one or more steps nor permuting the order of the steps upsets the stationary distribution of an MH within Gibbs sampler. The final phase involves removing redundant updates. Because MH steps generally depend on the current draws of {\it all} of the components of $\psi$ not marginalized out in that step, there are fewer redundant draws when some steps involve MH. Nonetheless, any redundant updates that are identified can safely be removed  in the trimming phase---by definition they do not affect the transition kernel. 
{\it The critical point is that unlike with an ordinary Gibbs sampler, we cannot simply replace some of the component draws of a PCG sampler with MH updates. Rather we must construct an MH within PCG sampler by applying the three-phase framework.}

Now suppose we wish to reduce the conditioning in an MH step. In Sampler Fragment~1, for example, if $p(\psi_3 | \psi_1, \psi_2)$ is a standard distribution with known normalization, then we can evaluate $p(\psi_2 | \psi_1) \propto p(\psi_1, \psi_2) = p(\psi_1, \psi_2, \psi_3) / p(\psi_3 | \psi_1, \psi_2)$ and sample $\psi_2 \sim  {\mathcal M}_{2|1}(\psi_2|\psi_1^\prime,\psi_2^\prime)$. Replacing Step $K+1$ of Sampler Fragment~1 with this reduced MH step, however, can alter the chain's stationary distribution in unpredictable ways. Instead, we propose to replace the full MH step with the reduced MH step {\it followed immediately} by a direct draw from the complete conditional of the reduced quantities. In Sampler Fragment~1 this would entail replacing Step $K+1$ with
\begin{description}
\itemsep=0in
\item[Step $K+1$ with Reduced Conditioning:] $\psi_2^{(t+1)} \sim {\mathcal M}_{2|1}(\psi_2|\psi_1^{(t+1)},\psi_2^{*})$
  and  
$\psi_3 \sim p(\psi_3 | \psi_1^{(t+1)},   \psi_2^{(t+1)})$.
\end{description}
This strategy ensures that the target stationary distribution is maintained. 
The expectation is that the updates of the reduced quantities will be trimmed after the steps are appropriately permuted and that the reduced MH step can be employed in the final sampler.  We denote the transition kernel of the full step (i.e., the reduced MH step followed by the complete conditional of the reduced quantities) by ${\cal M}^\star$. In Sampler Fragment~1, we rewrite the step with reduced conditioning 
\begin{description}
\itemsep=0in
\item[Step $K+1$ with Reduced Conditioning:] $(\psi_2^{(t+1)},\psi_3) \sim {\mathcal M}^\star_{2,3|1}(\psi_2, \psi_3|\psi_1^{(t+1)},\psi_2^{*}).$
\end{description}
Notice that this full update is not formally a MH update and has the advantage that it does not depend on all of the components of $\psi$. Thus, this step can follow a step that reduces $\psi_3$ out.

We now illustrate the construction of  a proper MH within PCG sampler for the spectral model given in~(\ref{eq:sesa}). For simplicity, we assume that $X$ is observed directly and we can ignore {(\ref{eq:nusesa})--(\ref{eq:sa})}. Figure~\ref{fig:spectralmodel61}(a) gives a six-step Gibbs sampler. Three of its steps require MH updates; the details of all the steps are given in Appendix B. The conditioning in four steps is reduced in Figure~\ref{fig:spectralmodel61}(b), and the steps are permuted in Figure~\ref{fig:spectralmodel61}(c) to allow the redundant draws of $X_L^{\star}$ and $\alpha^{\star}$ to be trimmed in four steps. Sampler 6, the resulting proper MH within PCG sampler, appears in Figure \ref{fig:spectralmodel67}. 

\begin{figure}[t]%
\centering
\subfiguretopcaptrue
\subfigure{%
\fbox{
\begin{minipage}[c][5cm][c]{0.45\linewidth}
 \begin{center}
\centerline{\bf Sampler 6}
\vskip -0.1in
\begin{steps}
\small
\itemsep=0in
\step $\mu\sim{\mathcal{M}_{\mu|X,\beta,\gamma,\phi}(\mu|\beta^\prime,\gamma^\prime,\mu^\prime,\phi^\prime)}$,
\step $\phi \sim{\mathcal{M}_{\phi|X,\beta,\gamma,\mu}(\phi|\beta^\prime,\gamma^\prime,\mu,\phi^\prime)}$,
\step $\beta \sim{\mathcal{M}_{\beta|X,\gamma,\mu,\phi}(\beta|\beta^\prime,\gamma^\prime,\mu,\phi)}$,
\step $\alpha \sim{p(\alpha|X,\beta,\gamma^\prime,\mu,\phi)}$,
\step $X_{L}\sim{p(X_{L}|X,\alpha,\beta,\gamma^\prime,\mu,\phi)}$,
\step $\gamma\sim{p(\gamma|X,X_{L},\alpha,\beta,\mu,\phi)}$.
\end{steps}
\end{center}
\end{minipage}
}}
\hspace{2pt}
\subfigure{%
\fbox{
\begin{minipage}[c][5cm][c]{0.47\linewidth}
 \begin{center}
\vskip -0.3in
\centerline{\bf Sampler 7}
\vskip -0.1in
\begin{steps}
\small
\itemsep=0in
\step $\mu\sim{\mathcal{M}_{\mu|X,\beta,\gamma,\phi}(\mu|\beta^\prime,\gamma^\prime,\mu^\prime,\phi^\prime)}$,
\step $\phi \sim{\mathcal{M}_{\phi|X,\beta,\gamma,\mu}(\phi|\beta^\prime,\gamma^\prime,\mu,\phi^\prime)}$,
\step $(\alpha,\beta)\sim{\mathcal{M}_{\alpha,\beta|X,\gamma,\mu,\phi}(\alpha,\beta|\alpha^\prime,\beta^\prime,\gamma^\prime,\mu,\phi)}$,
\step $X_{L}\sim{p(X_{L}|X,\alpha,\beta,\gamma^\prime,\mu,\phi)}$,
\step $\gamma\sim{p(\gamma|X,X_{L},\alpha,\beta,\mu,\phi)}$.
\end{steps}
\end{center}
\end{minipage}
}}%
\caption{Samplers 6 and 7. Figure~\ref{fig:spectralmodel61} verifies the propriety of Sampler 6, an MH within PCG sampler for fitting the spectral model in~(\ref{eq:sesa}). Sampler 7 blocks Steps 3 and 4 of Sampler 6 into a single MH step. Unfortunately, this results in an improper sampler, see Section \ref{sec:bnb}.}%
\label{fig:spectralmodel67}%
\end{figure}

\subsection{Using MH following a reduced step}
\label{sec:mhrs}

Using a full MH step immediately following a reduced step can be problematic. Sampler~5 illustrates this in its simplest form: a draw from a marginal distribution followed by an MH update of the conditional distribution of the remaining unknowns. As noted in Section~\ref{sec:cespcg} this is a particularly common problem in practice, even in its simplest form. In more complicated PCG samplers, the general phenomenon of introducing a full MH step immediately following a reduced step is the typical path by which introducing MH leads to an improper sampler. This is illustrated in Sampler~Fragment~1, where we are unable to replace the update in Step~$K$ with the reduced step $\psi_1^{(t+1)} \sim p(\psi_1|\psi_3^\prime)$. Thus, this case is particularly important and we propose two alternate samplers that maintain the basic structure of the underlying PCG sampler while allowing a form of MH in the step following a reduced step. Both solutions are conceptually straightforward. 

We begin by studying a special case that is useful for illustrating the two alternative samplers that we propose. We discuss the more general situation below. In particular we start in the general setting of Sampler~Fragment~1, but consider a PCG sampler in which $\psi_1$ is updated in Step~$K$ via a direct draw from the conditional distribution $p(\psi_1|\psi_3)$ of the marginal distribution $p(\psi_1,\psi_3)$, i.e., a reduced step. Again suppose that an MH update is required to update $\psi_2$ in Step~$K+1$. That is, Steps~$K$ and $K+1$ of the parent PCG sampler are
\begin{description}
\itemsep=0in
\item[Step $K$:] $\psi_1^{(t+1)} \sim p(\psi_1|\psi_3^\prime)$,\hfill(Sampler Fragment 2)
\item[Step $K+1$:] $\psi_2^{(t+1)} \sim p(\psi_2|\psi_1^{(t+1)},\psi_3^\prime)$.
\end{description}
\noindent Because MH is needed for Step $K+1$, these steps cannot be blocked. 

One straightforward general solution to the intractability of $p(\psi_2|\psi_1^{(t+1)},\psi_3^\prime)$ is simply to iterate the MH update within  Step~$K+1$ to obtain a draw from the conditional distribution,

\noindent {\it Iterated MH Strategy}:
\begin{description}
\itemsep=0in
\item[Step $K$:] $\psi_1^{(t+1)} \sim p(\psi_1|\psi_3^\prime)$,\hfill(Sampler Fragment 3)
\item[Step $K+1$:] Sample $\psi_2^{(t+l/L)} \sim {\mathcal M}_{2|1,3}(\psi_2|\psi_1^{(t+1)},\psi_2^{(t+(l-1)/L)},\psi_3^\prime)$, for $l=1,\dots,L$, to obtain $\psi_2^{(t+1)}\stackrel{\mbox{\tiny{approx}}}{\sim}  p(\psi_2|\psi_1^{(t+1)},\psi_3^\prime)$ at the subiteration $l=L$.
\end{description}
\noindent {We discuss methods for determining how large $L$ must be in {Sections~\ref{sec:cijmh} and~\ref{sec:smhpcg}}. With sufficiently large $L$, the Iterative MH Strategy delivers a draw that approximately follows $p(\psi_2|\psi_1^{(t+1)},\psi_3^\prime)$ and thus the sampler is {\it approximately proper}. In this special case the iterated MH strategy effectively blocks Steps~$K$ and $K+1$ to (nearly) deliver an independent draw from $p(\psi_1,\psi_2|\psi_3^\prime)$.}

Another solution to the intractability of $p(\psi_2|\psi_1^{(t+1)},\psi_3^\prime)$ is a joint MH update on the blocked version of Steps~$K$ and $K+1$,
\smallskip

\noindent {\it Joint MH Strategy}:
\begin{description}
\itemsep=0in
\item[Step $K$:] Update $(\psi_1,\psi_2)$ jointly via the MH jumping rule ${\mathcal J}_{1,2|3}(\psi_1,\psi_2|\psi_2^{(t)},\psi_3^\prime)=p(\psi_1|\psi_3^\prime)\linebreak {\mathcal J}_{2|1,3}(\psi_2|\psi_1,\psi_2^{(t)},\psi_3^\prime)$,
\item[Step $K+1$:] Omit.\hfill(Sampler Fragment 4)
\end{description}
\noindent The jumping rule in Step~$K$ of Sampler~Fragment~4 is exactly the concatenation of Step~$K$ and the jumping rule in Step~$K+1$ of Sampler~Fragment~3. By concatenating we avoid iteration.

The iterated MH strategy is in some sense a thinned version of the joint MH strategy. This, however, is an over simplification for two reasons. First, the iterated MH strategy updates $\psi_1$ only once for every $L$ updates of $\psi_2$ whereas the joint MH strategy updates both together. Second, although the jumping rule in the joint MH strategy is the same as that used by the iterated MH strategy at its first subiteration, the acceptance probabilities differ. This results in a systematic difference in the performance of the resulting samplers, see Section~\ref{sec:cijmh}.

Generalizing Sampler~Fragment~2,  Steps~$K$ and $K+1$ may not block even without MH. Suppose $\psi=(\psi_1,\psi_2,\psi_3,\psi_4)$ and the parent PCG sampler contains the two steps
\begin{description}
\itemsep=0in
\item[Step $K$:] $\psi_1^{(t+1)} \sim p(\psi_1|\psi_3^{(t)},\psi_4^\prime)$,\hfill(Sampler Fragment 5) 
\item[Step $K+1$:] $(\psi_2^{(t+1)},\psi_3^{(t+1)}) \sim p(\psi_2,\psi_3|\psi_1^{(t+1)},\psi_4^\prime)$, 
\end{description}
\noindent where Step~$K$ is a reduced step and Step~$K+1$ cannot be sampled directly. Here the conditional distributions cannot be blocked into a single step. {We can still use the iterated MH strategy in Step~$K+1$ to obtain a draw approximately from $p(\psi_2,\psi_3|\psi_1^{(t+1)},\psi_4^\prime)$ and an approximately proper sampler.} Likewise we can implement the joint MH strategy, using the jumping rule $p(\psi_1|\psi_3^{(t)},\psi_4^\prime){\mathcal J}_{2,3|1,4}(\psi_2,\psi_3|\psi_1,\psi_2^{(t)},\psi_3^{(t)},\psi_4^\prime)$. The stationary distribution of the joint jumping rule is $p(\psi_1|\psi_3^{(t)},\psi_4^\prime)p(\psi_2,\psi_3|\psi_1,\psi_4^\prime)$. Although a legitimate joint distribution on $(\psi_1,\psi_2,\psi_3)$, this does not correspond to a conditional distribution of $p(\psi)$.             
    
\subsection{To block or not to block}
\label{sec:bnb}

\begin{figure}[t]
\spacingset{1}
\begin{center}
    \includegraphics[height=2.2in,width=2.2in]{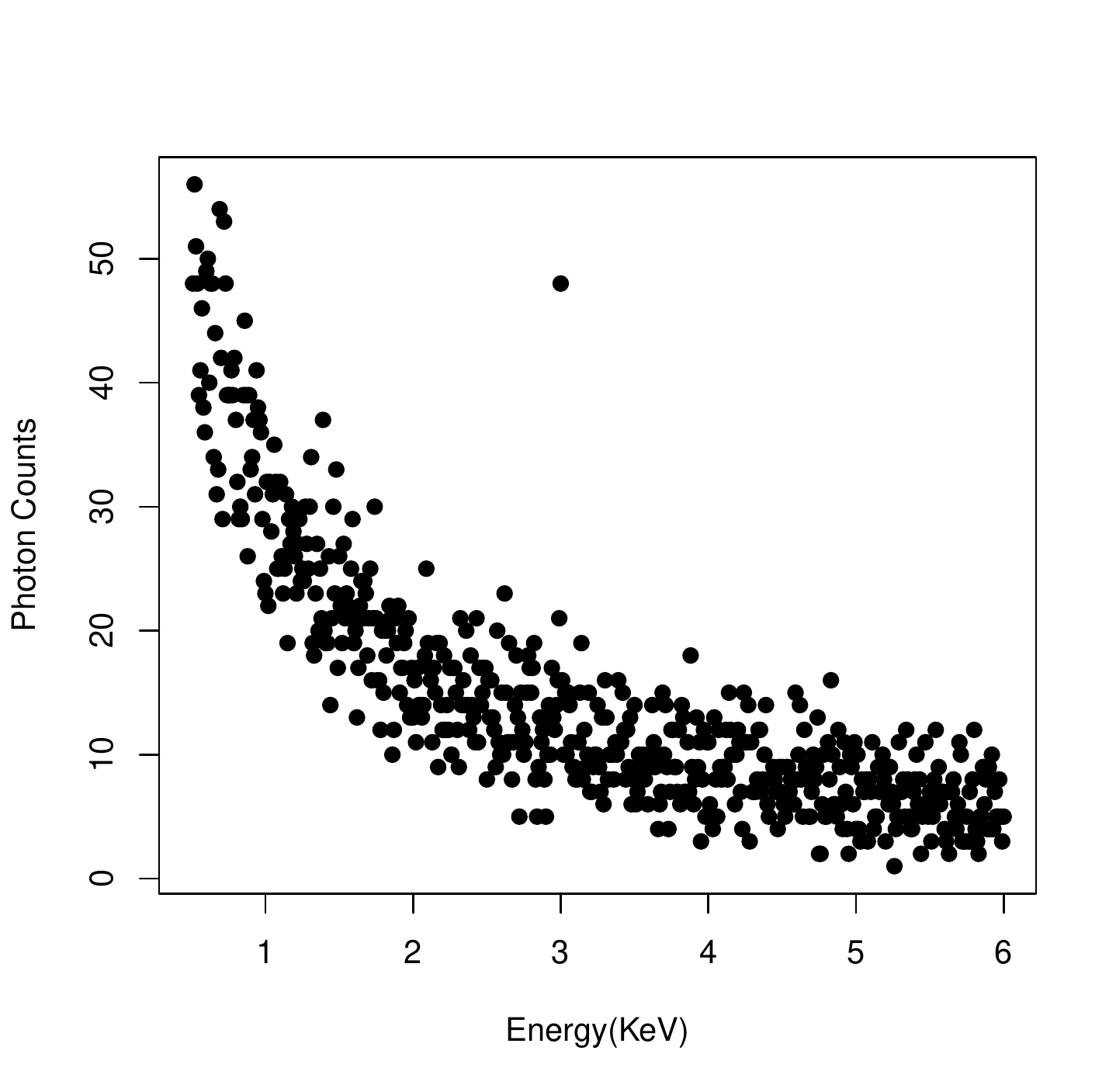}
    \caption{A dataset simulated under the spectral model~(\ref{eq:sesa}) and used in the simulation study in Section \ref{sec:bnb}.}
    \label{fig:simulate}
\end{center}
\end{figure}

Section~\ref{sec:mhrs} discusses the case where Step~$K+1$ of Sampler Fragment 2 requires MH. We now consider the case where Step~$K$ requires MH. In particular, 
\begin{description}
\itemsep=0in
\item[Step $K$:] $\psi_1^{(t+1)} \sim {\mathcal M}_{1|3}(\psi_1|\psi_1^{(t)},\psi_3^\prime)$,\hfill(Sampler Fragment 6) 
\item[Step $K+1$:] $\psi_2^{(t+1)} \sim p(\psi_2|\psi_1^{(t+1)},\psi_3^\prime)$. 
\end{description}
\noindent Sampler~Fragment~6 does not lead to convergence problems 
because the inputs to Step $K+1$ follow the correct distribution;  Figure~\ref{fig:ex3} verifies the stationary distribution of its parent chain. 

We might consider blocking the two steps in Sampler~Fragment~6 into a single MH update as
\begin{description}
\itemsep=0in
\item[Step $K$:] Update $(\psi_1,\psi_2)$ jointly via the MH jumping rule ${\mathcal J}_{1,2|3}(\psi_1,\psi_2|\psi_1^{(t)},\psi_2^{(t)},\psi_3^\prime)=\linebreak {\mathcal J}_{1|3}(\psi_1|\psi_1^{(t)},\psi_3^\prime)p(\psi_2|\psi_1,\psi_3^\prime)$,
\item[Step $K+1$:] Omit.\hfill(Sampler Fragment 7) 
\end{description}
\noindent The jumping rule in Sampler~Fragment~7 is exactly the concatenation of the jumping rules in the two steps of Sampler~Fragment~6. There is a fundamental difference, however, in that the concatenated jumping rule depends on $\psi_2^{(t)}$: if the MH proposal is rejected, $(\psi_1^{(t+1)},\psi_2^{(t+1)})=(\psi_1^{(t)},\psi_2^{(t)})$, whereas neither of the steps in Sampler~Fragment~6 depends on $\psi_2^{(t)}$. This means that care must be taken to ensure blocking in this way does not upset the stationary distribution of the chain. 

Steps 3 and 4 of Sampler 6 are an example of Sampler~Fragment~6, with $\psi_1=\beta$, $\psi_2=\alpha$ and $\psi_3=(\gamma,\mu,\phi)$. Blocking Steps 3 and 4 of Sampler 6 results in Sampler 7, see the second panel of Figure \ref{fig:spectralmodel67}. Unfortunately, this is an improper sampler, which we verify using a simulation study. We begin by generating an artificial data set consisting of $n=550$ bins with $\alpha=37.62$, $\beta=1$, $\gamma=40/37.62$, $\mu=250$, and $\phi=0.2$, see Figure~\ref{fig:simulate}. We run two versions of Sampler 7. Sampler 7(a) uses the concatenated jumping rule given in Sampler Fragment 7 to update $(\alpha,\beta)$, while Sampler 7(b) uses an independent bivariate normal jumping rule centered at the current value of $(\alpha,\beta)$. We use a uniform prior distribution for each parameter, and run 30,000 iterations of Samplers~6, 7(a), and 7(b) using the same starting values ($\alpha=30$, $\beta=3$, $\gamma=1$, $\mu=10$ and $\phi=0.5$). Scatter plots of $(\alpha,\beta,\phi)$ for the last 10,000 draws from the three samplers appear in Figure~\ref{fig:scatter1}, 
which shows that Samplers 7(a) and 7(b) underestimate the correlations of the target distribution; this effect is especially dramatic for Sampler 7(b). Figure~\ref{fig:specqq} compares the marginal distributions of $\alpha$, $\beta$, and $\phi$ generated with Samplers 6 and 7(b), and shows that Sampler 7(b) underestimates the marginal variances of all three parameters. (The marginals generated with Sampler 7(a) are more similar to those generated with Sampler 6.)

\begin{figure}[t]
\centering
\subfigure{%
\begin{minipage}{0.31\linewidth}
\begin{center}
  \includegraphics[width=2.4in,height=2.5in]{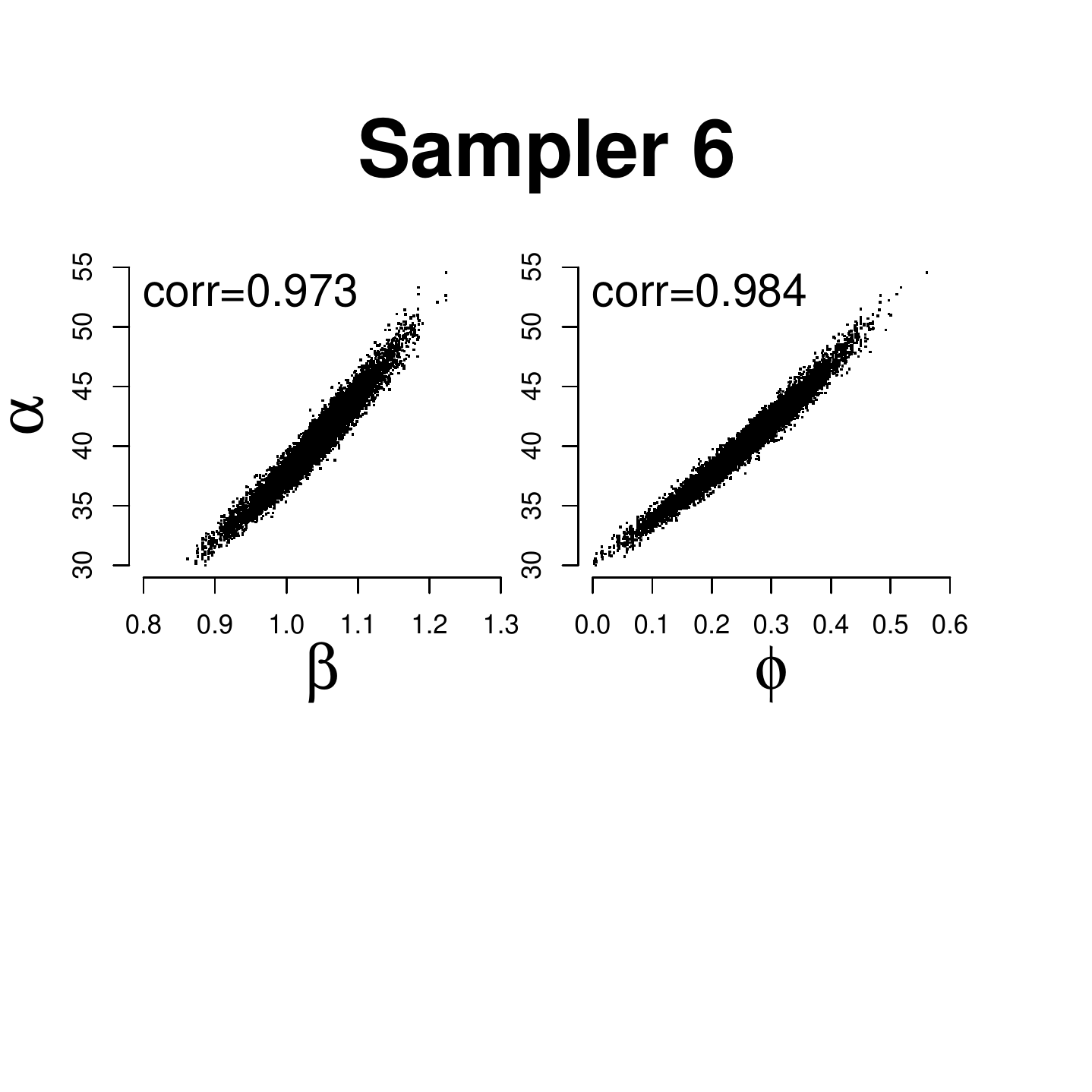}
\end{center}
\end{minipage}
}%
\hspace{8.5pt}%
\subfigure
{%
\begin{minipage}{0.31\linewidth}
\begin{center}
  \includegraphics[width=2.4in,height=2.5in]{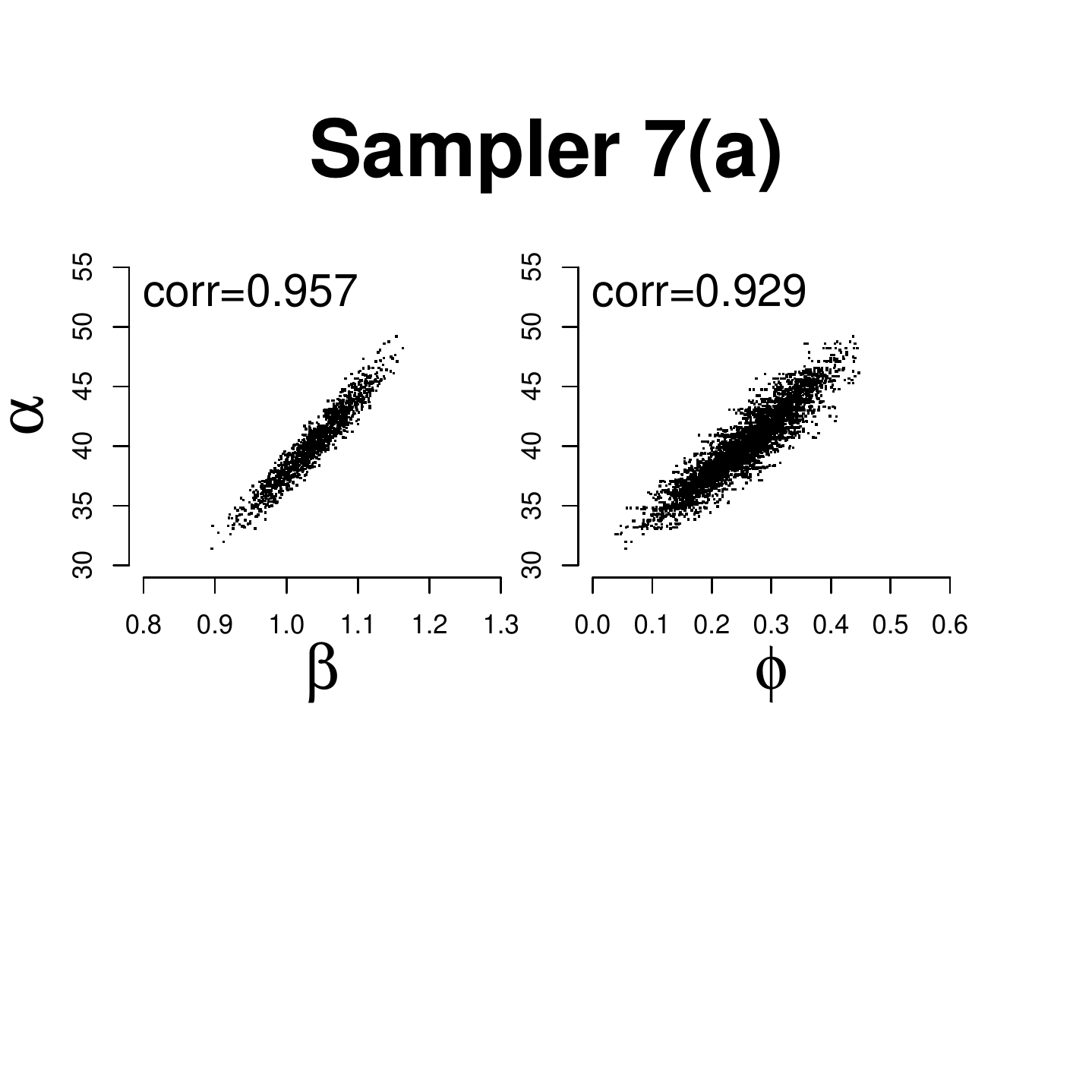}
\end{center}
\end{minipage}
}%
\hspace{8.5pt}%
\subfigure
{%
\begin{minipage}{0.31\linewidth}
\begin{center}
  \includegraphics[width=2.4in,height=2.5in]{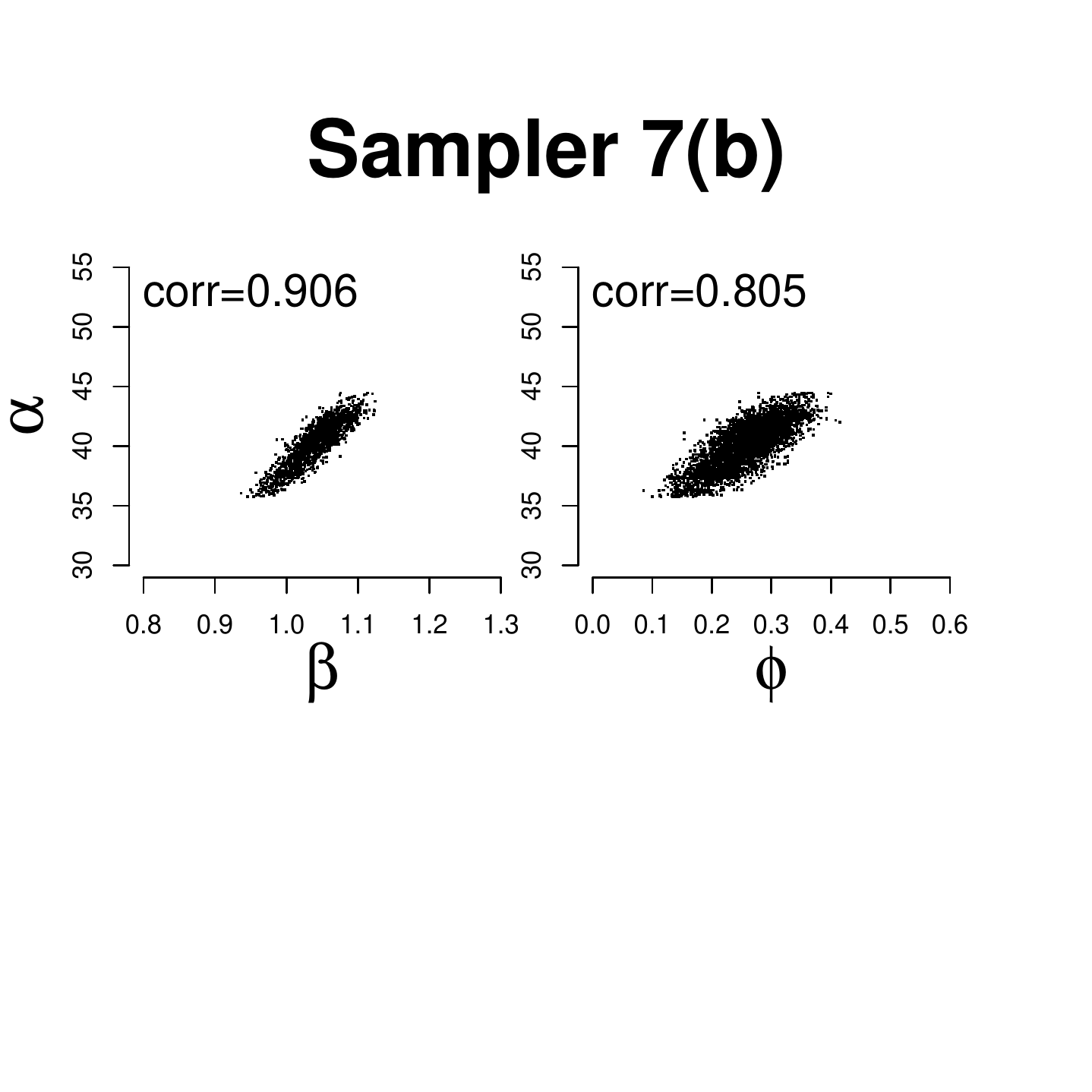}
\end{center}
\end{minipage}
}%
\vskip -1in
\caption{Scatter plots of $\alpha$, $\beta$ and $\phi$ for 10,000 draws from Samplers 6, 7(a) and 7(b) respectively. The two versions of Sampler 7 block the two steps of Sampler 6 that update $\alpha$ and $\beta$. Unfortunately, this results in an improper sampler. When updating $(\alpha,\beta)$, Sampler 7(a) uses the concatenation of Sampler 6's jumping rules for $\alpha$ and $\beta$, while Sampler 7(b) uses an independent bivariate normal jumping rule. The impropriety of Sampler 7(b) is especially dramatic.}
\label{fig:scatter1}
\end{figure} 

\begin{figure}[t]
\spacingset{1}
\begin{center}
  \includegraphics[width=5.5in]{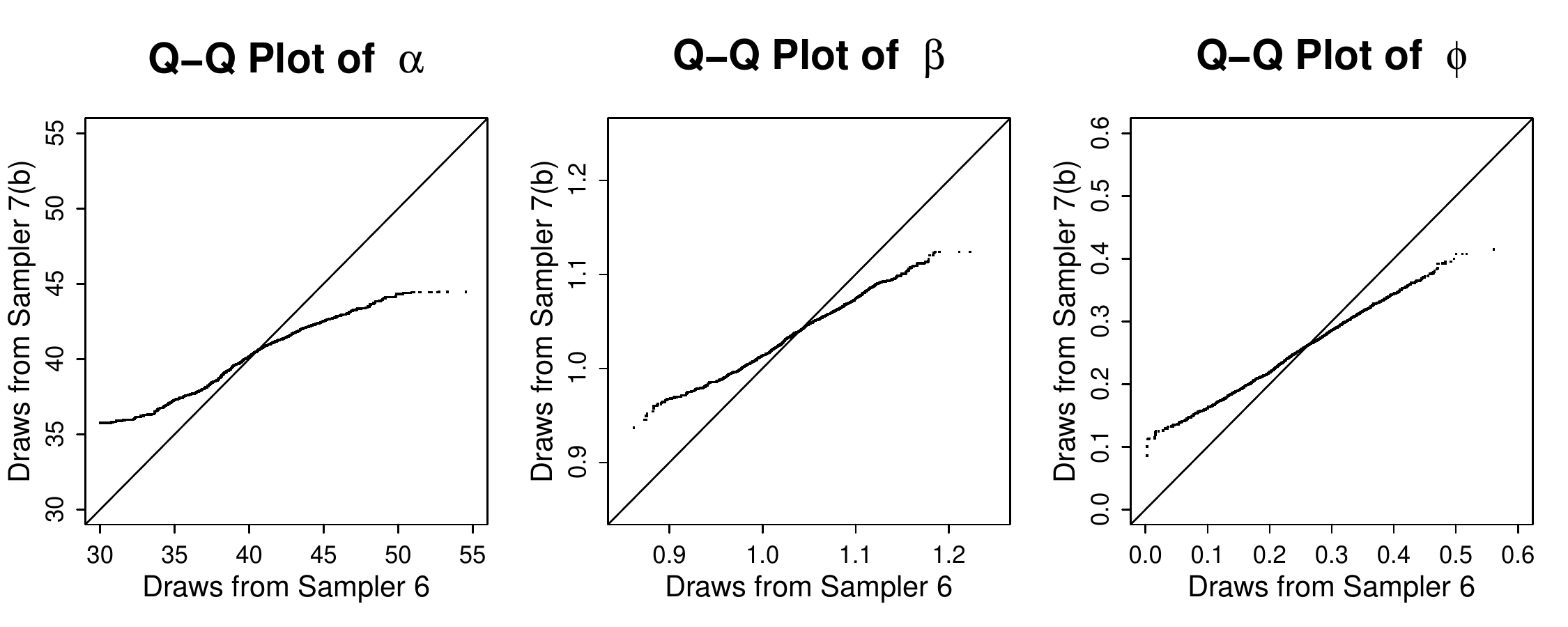}
    \caption{Quantile-quantile plots of $\alpha$, $\beta$ and $\phi$ corresponding to draws generated with Samplers 6 and 7(b). Sampler 7(b) severely underestimates the marginal variances of all three parameters.}
    \label{fig:specqq}
\end{center}
\end{figure}    

The problem with Sampler 7 can be understood in the terms of Section \ref{sec:mhrs}. Blocking the updates for $\alpha$ and $\beta$ results in an MH step that follows directly after a pair of reduced steps (the updates of $\mu$ and $\phi$). If $\mu$ and $\phi$ were known, and Steps 1 and 2 were removed, both versions of Samplers 7 would be proper. As it is, the stationary distribution of Sampler 7 cannot be verified with the three-phase framework. 

The comparison between Sampler~Fragments~6--7 is similar to that between the iterated and joint MH strategies in Section~\ref{sec:mhrs}. Theoretical perspectives on these choices appear in Section \ref{sec:the}.  



\section{Theory}
\label{sec:the}

\subsection{Comparing the iterated and joint MH strategies}
\label{sec:cijmh}

In this section we compare the iterated and joint MH strategies in terms of their acceptance probabilities. Although it is generally recognized that an acceptance probability of $20\%$ to $40\%$ is best for a symmetric Metropolis jumping rule~\citep{robe:etal:97}, we argue that the better choice between the two strategies is determined by maximizing the acceptance probability. This is because both the iterated and joint MH strategies start with the {\it same proposal}---they are numerically identical. The rule of thumb for tuning the acceptance probability to between $20\%$ and $40\%$ is based on comparing {\it different proposal distributions} with an eye on avoiding high acceptance rates because they typically correspond to jumping rules that propose very small steps. In this case the initial step sizes are the same and we aim to reduce correlation by increasing the jumping probability. We begin with theoretical results and then illustrate them numerically.

To simplify notation we suppress the conditioning on $\psi_3$ in Sampler~Fragments~3 and~4. This is equivalent to a formal comparison of the iterated and joint MH strategies as alternatives to the improper two-step Sampler~5. We assume that (i) the sampler has been verified to be proper so that $\pi=p$ and (ii) the jumping rule used to update $\psi_2$ does not depend on $\psi_1$, i.e., ${\mathcal J}_{2|1}(\psi_2|\psi_1^\prime,\psi_2^\prime)={\mathcal J}_{2|1}(\psi_2|\psi_2^\prime)$. While the transition kernel ${\mathcal{M}}_{2|1}({\psi}_{2}|{\psi}_{1}^{\prime},{\psi}_{2}^{\prime})$ will typically depend on ${\psi}_{1}^{\prime}$, the jumping rule often will not, for example, a symmetric Metropolis-type jumping rule does not.

The acceptance probability of the first draw in Step~$K+1$ of the iterated MH strategy is
\begin{eqnarray}
r_{\rm iter}=\frac{p(\psi_2^{\rm prop}|\psi_1^{(t+1/L)}){\mathcal J}_{2|1}(\psi_2^{(t)}|\psi_2^{\rm prop})}{p(\psi_2^{(t)}|\psi_1^{(t+1/L)}){\mathcal J}_{2|1}(\psi_2^{\rm prop}|\psi_2^{(t)})},
  \label{eq:itermh}
\end{eqnarray}    
where $\psi_1^{(t+1/L)} \sim p(\psi_1)$ and $\psi_2^{\rm prop} \sim {\mathcal J}_{2|1}(\psi_2|\psi_2^{(t)})$. With the joint MH strategy, it is 
\begin{eqnarray}
r_{\rm joint}=\frac{p(\psi_1^{\rm prop},\psi_2^{\rm prop}) \{p(\psi_1^{(t)}){\mathcal J}_{2|1}(\psi_2^{(t)}|\psi_2^{\rm prop})\}}{p(\psi_1^{(t)},\psi_2^{(t)}) \{p(\psi_1^{\rm prop}){\mathcal J}_{2|1}(\psi_2^{\rm prop}|\psi_2^{(t)})\}}=\frac{p(\psi_2^{\rm prop}|\psi_1^{\rm prop}) {\mathcal J}_{2|1}(\psi_2^{(t)}|\psi_2^{\rm prop})}{p(\psi_2^{(t)}|\psi_1^{(t)}) {\mathcal J}_{2|1}(\psi_2^{\rm prop}|\psi_2^{(t)})},
  \label{eq:jointmh}
\end{eqnarray} 
where $\psi_1^{\rm prop} \sim p(\psi_1)$ and $\psi_2^{\rm prop} \sim {\mathcal J}_{2|1}(\psi_2|\psi_2^{(t)})$. 
\begin{lemma} In the setting described in the previous paragraph,
\begin{eqnarray}
{\rm E}_{\pi}[r_{\rm iter}/r_{\rm joint}] \ge 1.
  \label{eq:eratio}
\end{eqnarray} 
\end{lemma}
The expectation in (\ref{eq:eratio}) is under the common stationary distribution, $\pi$, of both chains and is conditional on the random seed used at the start of each iteration. That is, since $(\psi_1^{(t+1/L)},\psi_2^{\rm prop})$ sampled under the iterated MH strategy and $(\psi_1^{\rm prop},\psi_2^{\rm prop})$ sampled under the joint MH strategy are drawn in exactly the same way, we assume these quantities are numerically equal. Expression~(\ref{eq:eratio}) asserts that while both strategies start with the same proposal---$(\psi_1^{(t+1/L)},\psi_2^{\rm prop})$ under the iterated MH strategy and $(\psi_1^{\rm prop},\psi_2^{\rm prop})$ under the joint---the iterated MH strategy is on average more likely to accept $\psi_2$. (The iterated MH strategy {\it always} accepts $\psi_1$.) 

\begin{proof} With the numerical equality of the proposals,
\begin{eqnarray}
\frac{r_{\rm iter}}{r_{\rm joint}}=\frac{p(\psi_2^{(t)}|\psi_1^{(t)})}{p(\psi_2^{(t)}|\psi_1^{(t+1/L)})},
  \label{eq:ratio}
\end{eqnarray} 
where $(\psi_1^{(t)},\psi_2^{(t)},\psi_1^{(t+1/L)}) \sim \pi(\psi_1^{(t)},\psi_2^{(t)})\pi_1(\psi_1^{(t+1/L)})$ with $\pi_1$ the $\psi_1$ marginal distribution of $\pi$. Because $(\psi_1^{(t)},\psi_2^{(t)}) \sim \pi$ and $\pi=p$, the numerator of~(\ref{eq:ratio}) is the conditional density of $\psi_2$ evaluated at $\psi_2^{(t)}$. This is not true of the denominator because $\psi_2^{(t)}$ is independent of $\psi_1^{(t+1/L)}$. Thus, we might expect that the numerator of~(\ref{eq:ratio}) is typically larger than the denominator, as claimed in~(\ref{eq:eratio}). 

Recalling that $\pi=p$, substituting~(\ref{eq:ratio}) into~(\ref{eq:eratio}), and applying Jensen's inequality, we need only verify that      
\begin{eqnarray}
\int {\rm log}\left[\pi(\psi_2|\psi_1)\right]\pi(\psi_1,\psi_2)d\psi_1 d\psi_2 \ge \int {\rm log}\left[\pi(\psi_2|\psi_1)\right]\pi(\psi_1)\pi(\psi_2)d\psi_1 d\psi_2. 
  \label{eq:jensen}
\end{eqnarray} 
Expression~(\ref{eq:jensen}) can be verified using a standard property of entropy along with the Kullback-Leiber (KL) divergence. In particular, because KL is nonnegative,
\begin{eqnarray}
\int {\rm log}\left[\pi(\psi_2)\right]\pi(\psi_1)\pi(\psi_2)d\psi_1 d\psi_2 \ge \int {\rm log}\left[\pi(\psi_2|\psi_1)\right]\pi(\psi_1)\pi(\psi_2)d\psi_1 d\psi_2. 
  \label{eq:kl}
\end{eqnarray} 
(The standard KL expression can be recovered by adding $\int {\rm log}\left[\pi(\psi_2)\right]\pi(\psi_1)\pi(\psi_2)d\psi_1 d\psi_2$ to both sides of~(\ref{eq:kl}).) But a standard property of entropy~\citep[e.g.,][]{ebra:etal:99} is 
\begin{eqnarray}
\int {\rm log}\left[\pi(\psi_2|\psi_1)\right]\pi(\psi_1,\psi_2)d\psi_1 d\psi_2 \ge \int {\rm log}\left[\pi(\psi_2)\right]\pi(\psi_1)\pi(\psi_2)d\psi_1 d\psi_2. 
  \label{eq:entropy}
\end{eqnarray}
Combining~(\ref{eq:kl}) and~(\ref{eq:entropy}) gives~(\ref{eq:jensen}) and hence the desired result. \QEDA
\end{proof}

\begin{figure}[t]
\spacingset{1}
\begin{center}
  \includegraphics[width=6.5in]{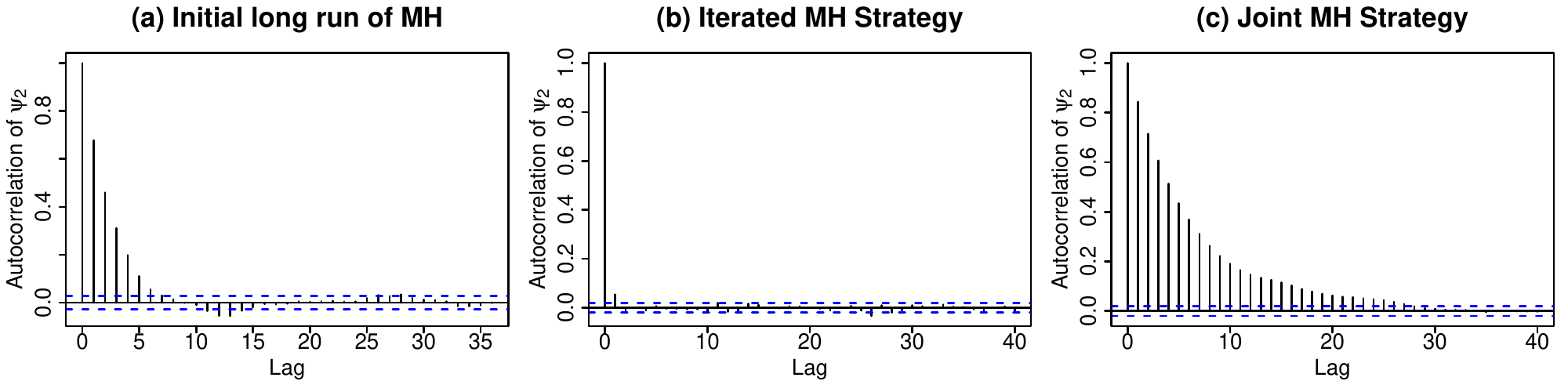}
    \caption{Autocorrelation functions of~$\psi_{2}$ for (a) an initial MH run of Step~2 of Sampler~5 with $\psi_1$ fixed, (b) the iterated MH strategy, and (c) the joint MH strategy, all under the bivariate normal simulation described in Section \ref{sec:cespcg}. Panel (a) shows that the initial MH runs deliver essentially independent draws after $7$ iterations, so that iterated MH strategy was run with $L=7$. Panels~(b) and~(c) show that the iterated strategy outperforms the joint one in terms of its computational efficiency.}
    \label{fig:auto}
\end{center}
\end{figure}  

We now return to the bivariate Gaussian simulation of Section~\ref{sec:cespcg}
to compare the computational performance of the iterated and joint MH strategies. Again we sample $\psi_1$ from its marginal distribution and use the same MH jumping rule to update $\psi_2$ according to its conditional distribution. The iterated strategy is run with $L=7$, in order to return $\psi_2^{(t+1)}$ that is essentially independent of $\psi_2^{(t)}$. {The value of $L$ was set using an initial MH run of $5,000$ iterations and inspecting the autocorrelation function.  The initial MH sampler delivers essentially independent draws after $7$ iterations, see Figure~\ref{fig:auto}(a). Of course, the computational cost per iteration of the iterated MH strategy depends on $L$. With $L=7$, each iteration requires eight univariate normal draws, whereas the joint strategy requires two. The autocorrelation functions of $\psi_{2}$ for both the iterated and joint MH strategies appear in Figure~\ref{fig:auto}(b)--(c) and show the clear computational advantage of the iterated MH strategy.} It returns essentially independent draws, whereas the joint MH strategy requires almost thirty iterations to obtain nearly independent draws.

In practice, it is important to check that the  value of $L$ used in Sampler Fragment~3 delivers samples that are essentially independent of the starting value of the iterated MH strategy. Fortunately, a simple diagnostic is available through the autocorrelation function of $\psi_2^{(t)}$ in Sampler Fragment~3, e.g., Figure~\ref{fig:auto}(b).  If the lag one autocorrelation is not essentially zero, the run should be repeated with a larger value of $L$. If $\psi_2$ is updated elsewhere in the sampler, the efficacy of the iterated MH strategy can be isolated by computing the correlation between the initial input of $\psi_2$ and the final output after iteration of the MH update in Step~$K+1$ of Sampler Fragment~3.

\subsection{Comparing the samplers with and without blocking}
\label{sec:tbnb}

To compare the blocking strategy in Sampler Fragment 7 with Sampler Fragment 6, we compute its acceptance rate, again suppressing the conditioning on $\psi_3$ for simplicity, as
\begin{eqnarray}
r_{\rm blocked}=\frac{p(\psi_1^{\rm prop},\psi_2^{\rm prop}){\mathcal J}_{1}(\psi_1^{(t)}|\psi_1^{\rm prop})p(\psi_2^{(t)}|\psi_1^{(t)})}{p(\psi_1^{(t)},\psi_2^{(t)}){\mathcal J}_{1}(\psi_1^{\rm prop}|\psi_1^{(t)})p(\psi_2^{\rm prop}|\psi_1^{\rm prop})}=\frac{p(\psi_1^{\rm prop}){\mathcal J}_{1}(\psi_1^{(t)}|\psi_1^{\rm prop})}{p(\psi_1^{(t)}){\mathcal J}_{1}(\psi_1^{\rm prop}|\psi_1^{(t)})}=r_{\rm not\;blocked},
  \label{eq:block}
\end{eqnarray} 
where $r_{\rm not\;blocked}$ is the acceptance probability of Step $K$ in Sampler Fragment 6, where there is no blocking. This means that Sampler Fragments 6 and 7 are identical in terms of their update of $\psi_1$, but whereas Sampler Fragment 6 updates $\psi_2$ with a new value at every iteration, blocking causes $\psi_2$ to only be updated if $\psi_1$ is updated. Thus, we expect the blocking strategy of Sampler Fragment 7 to reduce the efficiency of the sampler, and contrary to general advice regarding blocking \citep[e.g.,][]{liu:wong:kong:94}, the blocking strategy of Sampler Fragment 7 should be avoided. 

Together, the results of Sections~\ref{sec:cijmh} and~\ref{sec:tbnb} should be taken to discourage the combining of an MH update and a direct draw from a conditional distribution into a single MH update. 

\section{Examples}
\label{sec:exa}

\subsection{The simplest MH within PCG sampler}
\label{sec:smhpcg}

MH within PCG samplers are useful for fitting multi-component models in which part of the model must be fitted off-line. Consider a two-step sampler that updates $\psi_1$ and $\psi_2$ each in turn, but for computational reasons, we wish to update $\psi_1$ off-line. This may, for example, stem from the use of computer models that involve some costly evaluations in the update of $\psi_1$. As an illustration, we consider the problem of accounting for calibration uncertainty in high-energy astrophysics~\citep{lee:etal:11} using a special case of model~(\ref{eq:sa}) in Section~\ref{sec:saxa}:
\begin{eqnarray}
 Y_{j}{\sim}{\rm Poisson}\{A_{j}\alpha{E_{j}}^{-\beta}\},\mbox{ for }j=1,\dots,n.
  \label{eq:cali}
\end{eqnarray}
Here we consider the case where the effective area vector $A=(A_{1},\dots,A_{n})$ is not known, and must be estimated along with $\alpha$ and $\beta$. In-space calibration and sophisticated modelling of the instrument result in a representative sample of possible $A$ values.~\citet{lee:etal:11} shows how a Principal Component Analysis (PCA) of this sample can be used to derive a degenerate multivariate normal prior for $A$. In particular, we can write $A(Z)=A_0+QZ$, where $A_0\,(n\times 1)$ and $Q\,(n\times q)$ are known, the components of the $(q\times 1)$ vector, $Z$, are independent standard normal variables, and $q\ll n$. Since $A$ is a deterministic function of $Z$, we can confine attention to the parameter $(Z, \alpha, \beta)$. With the expectation that $Y$ would be relatively noninformative for $A(Z)$ and to simplify computation,~\citet{lee:etal:11} suggests adopting $p(Z)p(\alpha,\beta|Z,Y)$ as the target distribution for statistical inference, an approximation that they call {\it Pragmatic Bayes}. Thus, the target can be sampled by first drawing $Z\sim p(Z)$ and then updating $\alpha$ and $\beta$ given $Z$. Using a uniform prior for $\alpha$ and $\beta$: $p(\alpha,\beta)\propto 1$, the complete conditional for $\alpha$ is in closed form, but $\beta$ requires MH.

One might be tempted to implement the following improper MH within PCG sampler: 

{\begin{steps}
\itemsep=0in
\step $Z^{(t+1)}\sim{p(Z)}$, 
\hfill(Sampler 8)
\step $\beta^{(t+1)}\sim{\mathcal{M}_{\beta|Y,\alpha,A(Z)}(\beta|\alpha^{(t)},\beta^{(t)},A(Z^{(t+1)}))}$,
\step $\alpha^{(t+1)}\sim{p(\alpha|Y,\beta^{(t+1)},A(Z^{(t+1)}))}$.
\end{steps}}

\noindent This update of $\alpha$ and $\beta$ reflects the simple form of~(\ref{eq:cali}). Methods for fitting more general spectral models were considered by~\citet{lee:etal:11}. To derive an (approximately) proper sampler, we can remove the conditioning on $\alpha$ and implement the iterated MH strategy in Step~2:

{\begin{steps}
\itemsep=0in
\step $Z^{(t+1)}\sim{p(Z_{j})}$, 
\hfill(Sampler 9)
\step $\beta^{(t+l/L)}\sim{\mathcal{M}_{\beta|Y,A(Z)}(\beta|\beta^{(t+(l-1)/L)},A(Z^{(t+1)}))}$, for $l=1,\dots,L$,
\step $\alpha^{(t+1)}\sim{p(\alpha|Y,\beta^{(t+1)},A(Z^{(t+1)}))}$.
\end{steps}}

\noindent As suggested in Section~\ref{sec:cijmh}, we determine $L$ using an initial MH run of $1,000$ iterations and inspecting its autocorrelation function. We found that the component MH sampler delivers essentially independent draws of $\beta$ after $20$ iterations and thus set $L=20$ in Step 2 of Sampler~9.

We use a simulation study to illustrate the impropriety of Sampler 8. The data are simulated using $n=1078$ energy bins ranging from $0.225$ to $10.995$ keV, $q=7$, $Z_j=1.5\ (j=1,\dots,q)$, $\alpha=30$ and $\beta=1$. For each sampler, a chain of length 20,000 is run with a burnin of 10,000 from the starting values $Z=0$, $\alpha=1$ and $\beta=1$. Figure~\ref{fig:scacal} shows that using $L=20$ in Sampler~9 is sufficiently large and that Sampler 8 both underestimates the correlation of $Z_2$ and $\beta$ and the marginal variability of both $\alpha$ and (more dramatically) $\beta$. 

\begin{figure}[t]
\spacingset{1}
\begin{center}
  \includegraphics[width=6.2in]{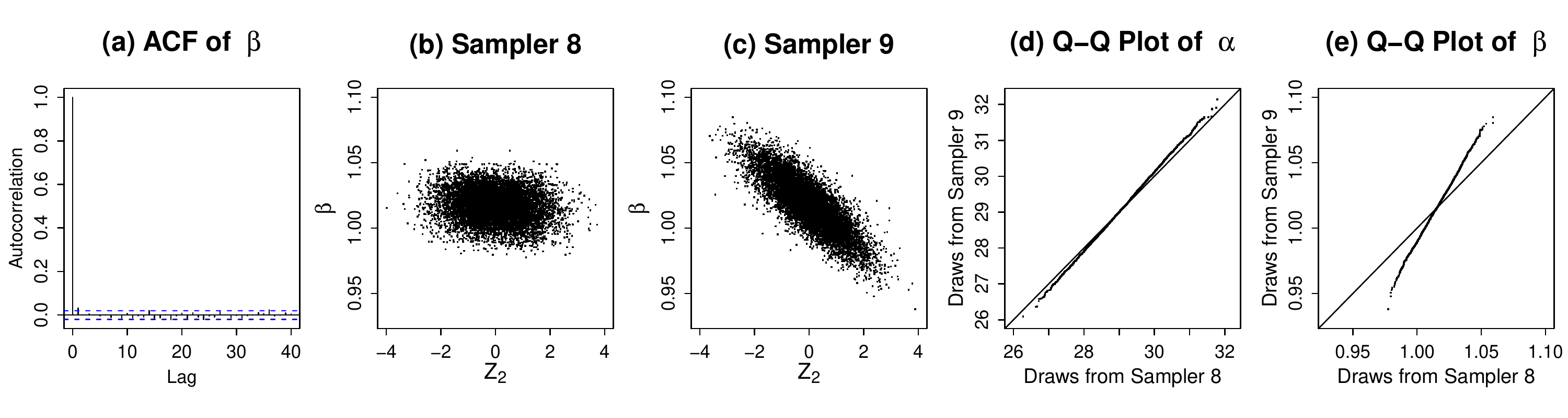}
    \caption{Numerical Evaluation of Samplers 8 and 9 using data simulated under model (\ref{eq:cali}).  (a): the diagnostic plot suggested in Section~\ref{sec:cijmh} for the choice of $L=20$ in Sampler 9. Since the lag-one autocorrelation of $\beta^{(t)}$ is essentially zero, $L$ is sufficiently large.  
(b) and (c): scatter plots of $Z_2$ and $\beta$ from Samplers 8 and 9 respectively.  (d) and (e): quantile-quantile plots of $\alpha$ and $\beta$ respectively. Sampler 9 is (approximately) proper while Sampler 8 is improper and underestimates the correlation between $Z_2$ and $\beta$ and also the marginal variability of both $\alpha$ and $\beta$.}
    \label{fig:scacal}
\end{center}
\end{figure} 

While~\citet{lee:etal:11} recognized the hazard of Sampler 8  and proposed Sampler 9, there are other examples in the literature where MH is used within a PCG sampler incorrectly, resulting in improper samplers. \citet{liu:etal:09}, for example, proposed a sampler very similar to Sampler 8 in structure, but in a completely different setting. To predict the temperature of a particular device at a certain time point, the parameters describing the physical properties of the device were linked to the other parameters via a computationally expensive computer model. One of the approaches described in \citet{liu:etal:09} for sampling all the model parameters from their posterior distribution was to update the physical-property parameters from their prior distributions first, and then sample the remaining parameters conditioning on the prior-generated values of the physical-property parameters. This approach was expected to reduce the confoundedness between the parameters and thus improve the mixture of the Markov chain. Since the updates of the other parameters relied on MH, this approach is problematic as illustrated in Section~\ref{sec:cespcg}. In analogy to Figure \ref{fig:scacal}, \citet{liu:etal:09} showed that the marginal distributions of the other parameters sampled via this approach were more variable than via the full Bayesian analysis or some other approaches. Other examples of improper samplers that are similar in structure to Sampler 8 were proposed in \citet{lunn:etal:09}, \citet{mccan:etal:10}, and even the popular WinBUGS package (Spiegelhalter, Thomas, Best and Lunn 2003), see \citet{wood:etal:12} for discussion. 
  
\subsection{Spectral analysis with narrow lines in high-energy astrophysics}
\label{sec:sanlha}

\begin{figure}[t]%
\centering
\subfiguretopcaptrue
\subfigure{%
\fbox{
\begin{minipage}[c][5cm][c]{0.47\linewidth}
 \begin{center}
\centerline{\bf Sampler 10}
\vskip -0.1in
\begin{steps}
\small
\itemsep=0in
\step $\mu \sim \mathcal{M}_{\mu|X,\alpha,\beta,\gamma,\phi,}{(\mu|\alpha^\prime,\beta^\prime,\gamma^\prime,\mu^\prime,\phi^\prime)}$,
\step $X_{L}\sim{p(X_{L}|X,\alpha^\prime,\beta^\prime,\gamma^\prime,\mu,\phi^\prime)}$,
\step $\alpha\sim p(\alpha|X,X_{L},\beta^\prime,\gamma^\prime,\mu,\phi^\prime)$,
\step $\beta\sim \mathcal{M}_{\beta|X,X_{L},\alpha,\gamma,\mu,\phi}(\beta|X_{L},\alpha,\beta^\prime,\gamma^\prime,\mu,\phi^\prime)$,
\step $\gamma\sim{p(\gamma|X,X_{L},\alpha,\beta,\mu,\phi)}$,
\step $\phi\sim{\mathcal{M}_{\phi|X,X_{L},\alpha,\beta,\gamma,\mu}(\phi|X_{L},\alpha,\beta,\gamma,\mu,\phi^\prime)}$.
\end{steps}
\end{center}
\end{minipage}
}}
\hspace{2pt}
\subfigure{%
\fbox{
\begin{minipage}[c][5cm][c]{0.45\linewidth}
 \begin{center}
\vskip -0.3in
\centerline{\bf Sampler 11}
\vskip -0.1in
 \begin{steps}
\small
\itemsep=0in
\step $\mu\sim{\mathcal{M}_{\mu|X,\beta,\gamma,\phi}(\mu|\beta^\prime,\gamma^\prime,\mu^\prime,\phi^\prime)}$,
\step $(\beta,\phi)\sim{\mathcal{M}_{\beta,\phi|X,\gamma,\mu}(\beta,\phi|\beta^\prime,\gamma^\prime,\mu,\phi^\prime)}$,
\step $\alpha\sim{p(\alpha|X,\beta,\gamma^\prime,\mu,\phi)}$,
\step $X_{L}\sim{p(X_{L}|X,\alpha,\beta,\gamma^\prime,\mu,\phi)}$,
\step $\gamma\sim{p(\gamma|X,X_{L},\alpha,\beta,\mu,\phi)}$.
\end{steps}
\end{center}
\end{minipage}
}}%
\caption{Samplers 10 and 11. Sampler 10 is the proper MH within PCG sampler for the spectral model~(\ref{eq:sesa}) with the lowest degree of partial collapsing, while Sampler 11 is that with the highest degree of partial collapsing.}%
\label{fig:standandproper}%
\end{figure} 

\begin{figure}[t]%
\centering
\subfiguretopcaptrue
\subfigure[Parent MH within Gibbs Sampler]{%
\framebox[0.995\width]{
\begin{minipage}[c][3.3cm][c]{0.485\linewidth}
\begin{steps}
\small
\itemsep=-0.1in
\step ${p(X_{L}|X,\alpha^\prime,\beta^\prime,\gamma^\prime,\mu^\prime,\phi^\prime)}$
\step $p(\alpha|X,X_{L},\beta^\prime,\gamma^\prime,\mu^\prime,\phi^\prime)$
\step ${\mathcal{M}_{\beta|X,X_{L},\alpha,\gamma,\mu,\phi}(\beta|X_{L},\alpha,\beta^\prime,\gamma^\prime,\mu^\prime,\phi^\prime)}$
\step ${p(\gamma|X,X_{L},\alpha,\beta,\mu^\prime,\phi^\prime)}$
\step ${\mathcal{M}_{\mu|X,X_{L},\alpha,\beta,\gamma,\phi}(\mu|X_{L},\alpha,\beta,\gamma,\mu^\prime,\phi^\prime)}$
\step ${\mathcal{M}_{\phi|X,X_{L},\alpha,\beta,\gamma,\mu}(\phi|X_{L},\alpha,\beta,\gamma,\mu,\phi^\prime)}$
\end{steps}
\end{minipage}
}}%
\hspace{0pt}%
\subfigure[Reduce Conditioning]{%
\framebox[0.995\width]{
\begin{minipage}[c][3.3cm][c]{0.485\textwidth}
 \begin{steps}
\small
\itemsep=-0.1in
\step ${p(X_{L}^\star|X,\alpha^\prime,\beta^\prime,\gamma^\prime,\mu^\prime,\phi^\prime)}$
\step $p(\alpha^\star,X_{L}^\star|X,\beta^\prime,\gamma^\prime,\mu^\prime,\phi^\prime)$
\step ${\mathcal{M}^\star_{\beta,X_{L},\alpha,\phi|X,\gamma,\mu}(\beta^\star,X_{L}^\star,\alpha^\star,\phi^\star|\beta^\prime,\gamma^\prime,\mu^\prime,\phi^\prime)}$
\step ${p(\gamma|X,X_{L}^\star,\alpha^\star,\beta^\star,\mu^\prime,\phi^\star)}$
\step ${\mathcal{M}^\star_{\mu,X_{L},\alpha|X,\beta,\gamma,\phi}(\mu,X_{L}^\star,\alpha^\star|\beta^\star,\gamma,\mu^\prime,\phi^\star)}$
\step ${\mathcal{M}^\star_{\phi,X_{L},\alpha,\beta|X,\gamma,\mu}(\phi,X_{L},\alpha,\beta|\beta^\star,\gamma,\mu,\phi^\star)}$

\end{steps}
\end{minipage}
}}\\
\subfigure[Permute]{%
\framebox[0.995\width]{
\begin{minipage}[c][3.3cm][c]{0.485\linewidth}
 \begin{steps}
\small
\itemsep=-0.1in
\step ${\mathcal{M}^\star_{\mu,X_{L},\alpha|X,\beta,\gamma,\phi}(\mu,X_{L}^\star,\alpha^\star|\beta^\prime,\gamma^\prime,\mu^\prime,\phi^\prime)}$
\step ${\mathcal{M}^\star_{\phi,X_{L},\alpha,\beta|X,\gamma,\mu}(\phi^\star,X_{L}^\star,\alpha^\star,\beta^\star|\beta^\prime,\gamma^\prime,\mu,\phi^\prime)}$
\step ${\mathcal{M}^\star_{\beta,X_{L},\alpha,\phi|X,\gamma,\mu}(\beta,X_{L}^\star,\alpha^\star,\phi|\beta^\star,\gamma^\prime,\mu,\phi^\star)}$
\step $p(\alpha,X_{L}^\star|X,\beta,\gamma^\prime,\mu,\phi)$
\step ${p(X_{L}|X,\alpha,\beta,\gamma^\prime,\mu,\phi)}$
\step ${p(\gamma|X,X_{L},\alpha,\beta,\mu,\phi)}$
\end{steps}
\end{minipage}
}}
\hspace{-3pt}%
\subfigure[Trim]{%
\framebox[0.995\width]{
\begin{minipage}[c][3.3cm][c]{0.485\linewidth}
\vskip -0.25in
\begin{steps}
\small
\itemsep=-0.1in
\step ${\mathcal{M}_{\mu|X,\beta,\gamma,\phi}(\mu|\beta^\prime,\gamma^\prime,\mu^\prime,\phi^\prime)}$
\step ${\mathcal{M}_{\beta,\phi|X,\gamma,\mu}(\beta,\phi|\beta^\prime,\gamma^\prime,\mu,\phi^\prime)}$
\step $p(\alpha|X,\beta,\gamma^\prime,\mu,\phi)$
\step ${p(X_{L}|X,\alpha,\beta,\gamma^\prime,\mu,\phi)}$
\step ${p(\gamma|X,X_{L},\alpha,\beta,\mu,\phi)}$
\end{steps}
\end{minipage}
}}
\caption{Three-phase framework used to derive Sampler~11 from its parent MH within Gibbs sampler. The parent sampler appears in (a). The conditioning in Steps 2, 3, 5, and 6 is reduced in (b) and the steps are permuted in (c) to allow redundant draws of $X_L^{\star}$, $\alpha^{\star}$, $\beta^{\star}$, and $\phi^{\star}$ to be trimmed in Steps 1--4. The resulting proper Sampler 11 appears in (d).}%
\label{fig:spectralmodel111}%
\end{figure}

\begin{figure}[t]
\centering
\subfigure{%
\begin{minipage}{0.51\linewidth}
\begin{center}
  \includegraphics[width=2.8in]{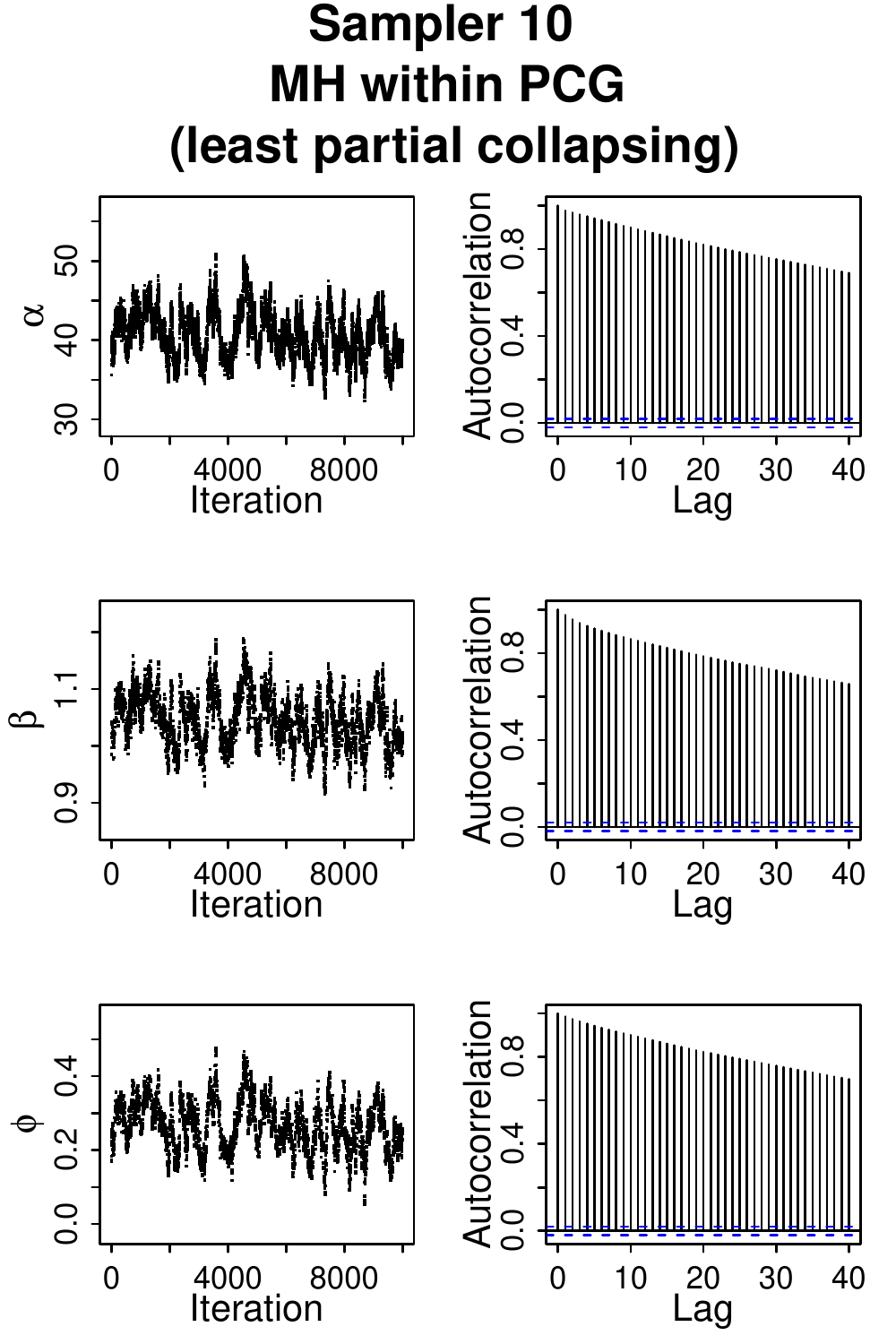}
\end{center}
\end{minipage}
}%
\hspace{-0.55in}%
\subfigure
{%
\begin{minipage}{0.51\linewidth}
\begin{center}
  \includegraphics[width=2.8in]{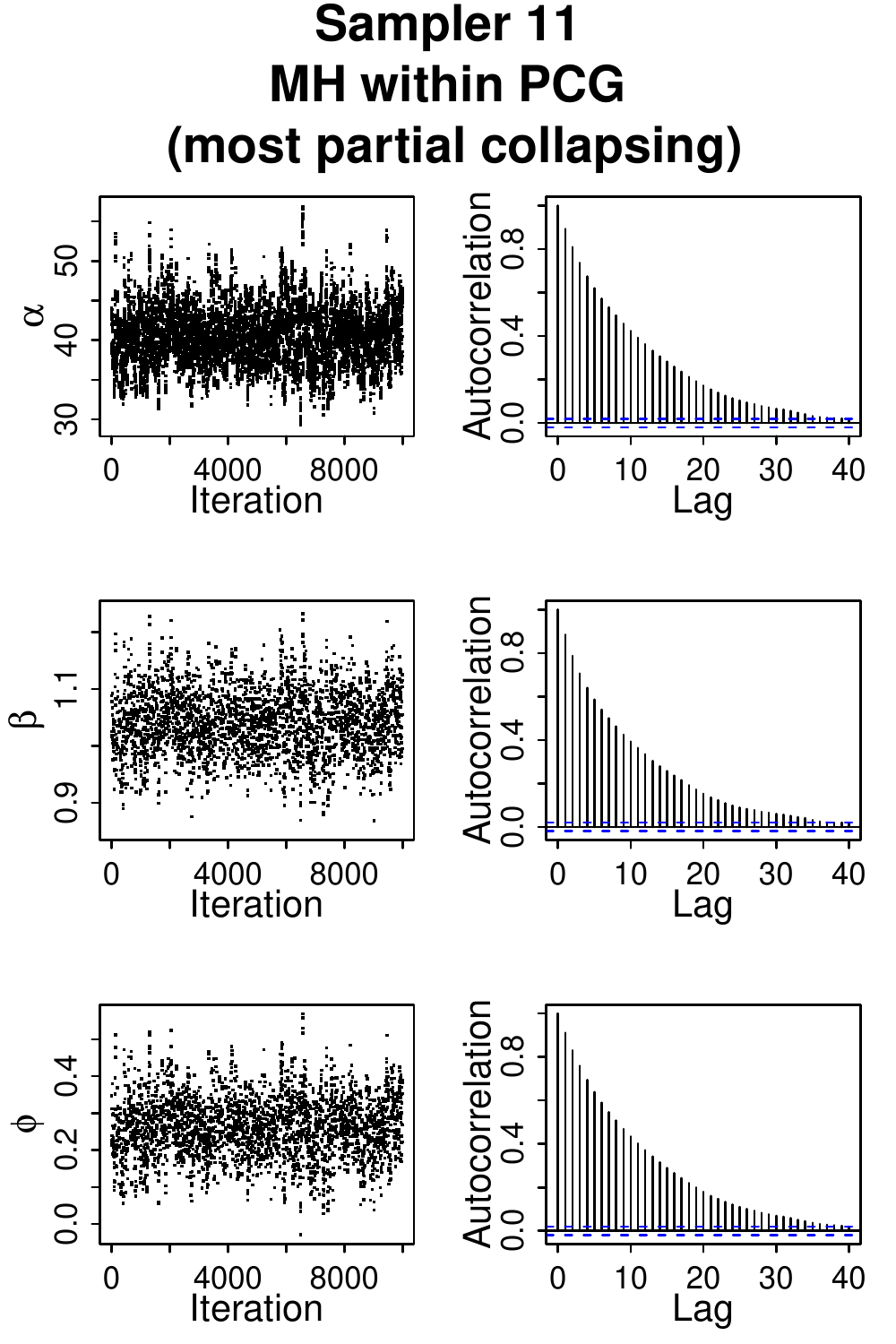}
\end{center}
\end{minipage}
}%
\caption{Comparing Samplers 10 and 11 using data simulated under model (\ref{eq:sesa}). The first two columns are the time-series and autocorrelation plots for the posterior draws of $\alpha$, $\beta$, and $\phi$ respectively from Sampler 10, while the last two columns are those from Sampler 11. Sampler 11 performs significantly better than Sampler 10.}
\label{fig:conspectral}
\end{figure}

\begin{figure}[t]%
\centering
\subfiguretopcaptrue
\subfigure{%
\fbox{
\begin{minipage}[c][2.7cm][c]{0.87\linewidth}
 \begin{center}
\centerline{\bf Sampler 12}
\vskip -0.1in
\begin{description}
\itemsep=0in
\small
\item[Step $1$:] $Z_{i}\sim{p(Z_{i}|Y,\beta^\prime,\Sigma^\prime)}$, for $i=1,\dots,100$,
\item[Step $2$:]  $\sigma_{j}^{2}\sim{p(\sigma_{j}^{2}|Y,Z,\beta^\prime)}$, for $j=1,\dots,5$,
\item[Step $3$:] $\beta_{j}\sim{p(\beta_{j}|Y,Z,\Sigma)}$, for $j=1,\dots,5$.
\end{description}
\end{center}
\end{minipage}
}}
\hspace{8pt}
\subfigure{%
\fbox{
\begin{minipage}[c][3.5cm][c]{0.87\linewidth}
 \begin{center}
\centerline{\bf Sampler 13}
\vskip -0.1in
\begin{description}
\itemsep=0in
\small
\item[Step $1$:] $\sigma_{1}^{2}\sim{p(\sigma_{1}^{2}|Y,Z^\prime,\beta^\prime)}$,
\item[Step $j$:] $\sigma_{j}^{2}\sim{\mathcal M}_{\sigma_{j}^2|Y,\beta,\sigma_{1}^{2},\dots,\sigma_{j-1}^{2},\sigma_{j+1}^{2},\dots,\sigma_{5}^{2}}(\sigma_{j}^2|\beta^\prime,\sigma_{1}^{2},\dots,\sigma_{j-1}^{2},{\sigma_{j}^{2}}^\prime,\dots,{\sigma_{5}^{2}}^\prime)$, for $j=2,\dots,5$, 
\item[Step $6$:] $Z_{i}\sim{p(Z_{i}|Y,\beta^\prime,\Sigma)}$, for $i=1,\dots,100$,
\item[Step $7$:] $\beta_{j}\sim{p(\beta_{j}|Y,Z,\Sigma)}$, for $j=1,\dots,5$.
\end{description}
\end{center}
\end{minipage}
}}%
\caption{Two samplers for fitting (\ref{eq:factor}). Sampler 12 is a standard Gibbs sampler and Sampler 13 is a proper MH within PCG sampler. Notice that Sampler 13 does not condition on $Z$ in its updates of $\sigma_2^2,\dots,\sigma_5^2$.}%
\label{fig:parentproper}%
\end{figure}

\begin{figure}[t]%
\centering
\subfiguretopcaptrue
\subfigure[Parent Gibbs Sampler (Sampler 12)]{%
\fbox{
\begin{minipage}[c][1.6cm][c]{0.98\linewidth}
\begin{description}
\itemsep=-0.1in
\item[Step $1$:]${p(Z_{i}|Y,\beta^\prime,\Sigma^\prime)}$, for $i=1,\dots,100$
\item[Step $2$:]${p(\sigma_{j}^{2}|Y,Z,\beta^\prime)}$, for $j=1,\dots,5$
\item[Step $3$:]${p(\beta_{j}|Y,Z,\Sigma)}$, for $j=1,\dots,5$
\end{description}
\end{minipage}
}}%
\vskip -2pt%
\subfigure[Reduce Conditioning]{%
\fbox{
\begin{minipage}[c][3cm][c]{0.98\linewidth}
 \begin{description}
\itemsep=-0.1in
\item[Step $1$:] ${p(Z_{i}^{*}|Y,\beta^\prime,\Sigma^\prime)}$, for $i=1,\dots,100$
\item[Step $2$:] ${p(\sigma_{1}^{2}|Y,Z^\star,\beta^\prime)}$
\item[Step $1+j$:]${\mathcal M}^\star_{\sigma_{j}^2,Z|Y,\beta,\sigma_{1}^{2},\dots,\sigma_{j-1}^{2},\sigma_{j+1}^{2},\dots,\sigma_{5}^{2}}(\sigma_{j}^2,Z^{*}|\beta^\prime,\sigma_{1}^{2},\dots,\sigma_{j-1}^{2},{\sigma_{j}^{2}}^\prime,\dots,{\sigma_{5}^{2}}^\prime)$, for $j=2,3,4$
\vskip 0.05in
\item[Step $6$:]${\mathcal M}^\star_{\sigma_{5}^2,Z|\beta,\sigma_{1}^{2},\dots,\sigma_{4}^{2}}(\sigma_{5}^2,Z|\beta^\prime,\sigma_{1}^{2},\dots,\sigma_{4}^{2},{\sigma_{5}^{2}}^\prime)$
\vskip 0.05in
\item[Step $7$:]${p(\beta_{j}|Y,Z,\Sigma)}$, for $j=1,\dots,5$
\end{description}
\end{minipage}
}}\\
\vskip -2pt
\subfigure[Permute]{%
\fbox{
\begin{minipage}[c][2.3cm][c]{0.98\linewidth}
 \begin{description}
\itemsep=-0.1in
\item[Step $1$:]${p(\sigma_{1}^{2}|Y,Z^\prime,\beta^\prime)}$
\item[Step $j$:]${\mathcal M}^\star_{\sigma_{j}^2,Z|Y,\beta,\sigma_{1}^{2},\dots,\sigma_{j-1}^{2},\sigma_{j+1}^{2},\dots,\sigma_{5}^{2}}(\sigma_{j}^2,Z^{*}|\beta^\prime,\sigma_{1}^{2},\dots,\sigma_{j-1}^{2},{\sigma_{j}^{2}}^\prime,\dots,{\sigma_{5}^{2}}^\prime)$, for $j=2,\dots,5$
\vskip 0.05in
\item[Step $6$:] ${p(Z_{i}|Y,\beta^\prime,\Sigma)}$, for $i=1,\dots,100$
\item[Step $7$:] ${p(\beta_{j}|Y,Z,\Sigma)}$, for $j=1,\dots,5$
\end{description}
\end{minipage}
}}
\vskip -2pt
\subfigure[Trim (Sampler 13)]{%
\fbox{
\begin{minipage}[c][2.3cm][c]{0.98\linewidth}
 \begin{description}
\itemsep=-0.1in
\item[Step $1$:] $p(\sigma_{1}^{2}|Y,Z^\prime,\beta^\prime)$,
\item[Step $j$:] ${\mathcal M}_{\sigma_{j}^2|Y,\beta,\sigma_{1}^{2},\dots,\sigma_{j-1}^{2},\sigma_{j+1}^{2},\dots,\sigma_{5}^{2}}(\sigma_{j}^2|\beta^\prime,\sigma_{1}^{2},\dots,\sigma_{j-1}^{2},{\sigma_{j}^{2}}^\prime,\dots,{\sigma_{5}^{2}}^\prime)$, for $j=2,\dots,5$, 
\item[Step $6$:] $p(Z_{i}|Y,\beta^\prime,\Sigma)$, for $i=1,\dots,100$,
\item[Step $7$:] $p(\beta_{j}|Y,Z,\Sigma)$, for $j=1,\dots,5$.
\end{description}
\end{minipage}
}}
\vskip -0.1in
\caption{Using the three-phase framework to derive Sampler~13 from its parent Gibbs sampler, i.e., Sampler 12. The parent Gibbs sampler is in (a); the conditioning in Steps 3--6 is reduced in (b); and the steps are permuted in (c) to allow redundant draws of $Z^{\star}$ to be trimmed in Steps 2--5. The resulting proper Sampler 13 is in (d).}%
\label{fig:factormodel}%
\end{figure}

\begin{figure}[t]
\centering
\subfigure{%
\begin{minipage}{0.5\linewidth}
\begin{center}
  \includegraphics[width=2.8in]{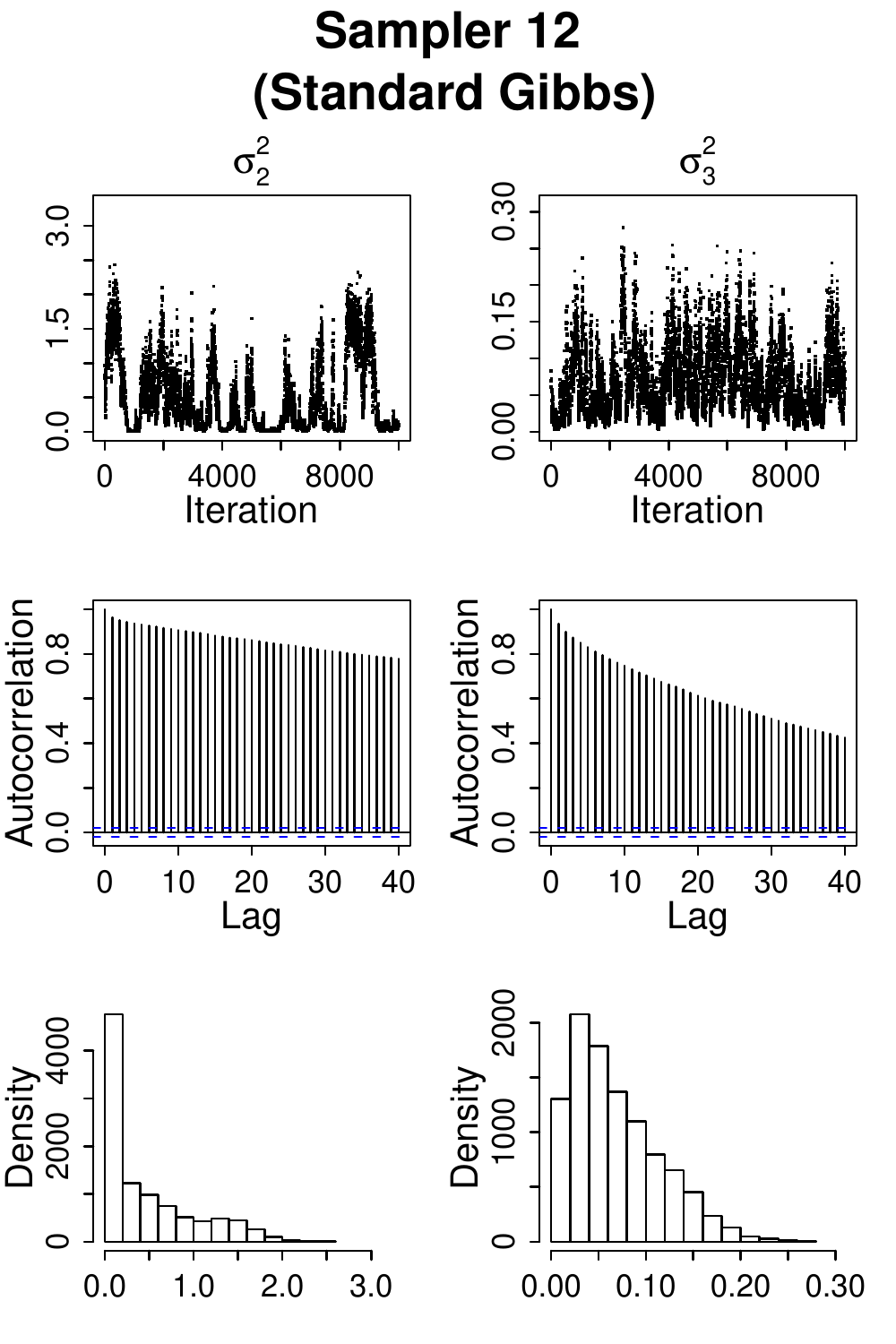}
\end{center}
\end{minipage}
}%
\hspace{-0.55in}%
\subfigure
{%
\begin{minipage}{0.5\linewidth}
\begin{center}
  \includegraphics[width=2.8in]{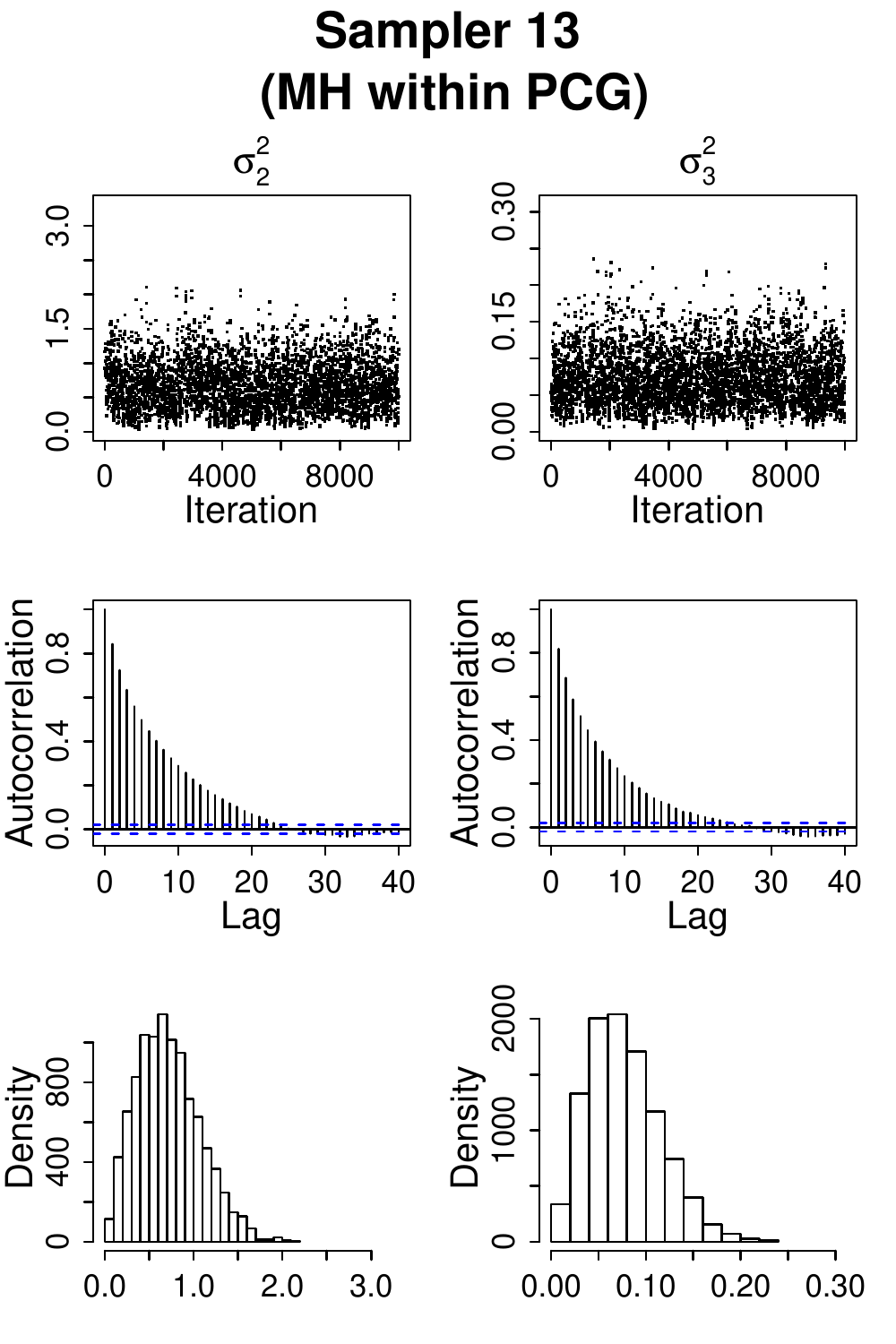}
\end{center}
\end{minipage}
}%
\caption{Comparing Samplers 12 and 13 using data simulated under the factor analysis model (\ref{eq:factor}). The first two columns are the time-series, autocorrelation, and histogram plots for the posterior draws of $\sigma_{2}^2$ and $\sigma_{3}^2$ respectively from Sampler 12, while the last two columns are those from Sampler 13. Sampler 13 performs significantly better than Sampler 12 both in terms of convergence properties and in its estimates of the marginal posterior distributions.}
\label{fig:refactor}
\end{figure}

Section~\ref{sec:bnb} uses a simulation study to illustrate a potential problem with Sampler Fragment 7, that is, how the blocking of an MH update and a direct draw from a conditional distribution can result in an improper sampler. Here we use the same simulation study to illustrate the improved convergence properties of three proper MH within PCG samplers relative to their parent Gibbs sampler. The only difference is that for each sampler here, a chain of 20,000 iterations is run with a burnin of 10,000 iterations. 

As pointed out in Section~\ref{sec:saxa}, the standard Gibbs sampler for the spectral model~(\ref{eq:sesa}) breaks down since the resulting subchain for $\mu$ does not move from its starting value~\citep{park:vand:09}. To solve this problem, we sample $\mu$ without conditioning on $X_{L}$ and obtain an MH within PCG sampler, i.e., Sampler~10, given in the first panel of Figure~\ref{fig:standandproper}. Sampler~6 in Figure \ref{fig:spectralmodel67} is another MH within PCG sampler but  with a higher degree of partial collapsing, by which we mean more quantities are marginalized out in Sampler~6 than in Sampler~10.
Not only does Sampler 6 update $\mu$ without conditioning on $X_{L}$, but it also marginalizes $\alpha$ out of its first three steps, whereas Sampler 10 does not remove $\alpha$ from any step. Sampler 11 attempts to further improve Sampler 6 by blocking the MH updates of $\beta$ and $\phi$, see the second panel of Figure~\ref{fig:standandproper}. Unlike Sampler 7 which also blocks 2 steps of Sampler 6, Sampler 11 is proper, see Figure \ref{fig:spectralmodel111}. Thus Samplers 6, 10 and 11 are all proper MH within PCG samplers with common parent Gibbs sampler given in Figure \ref{fig:spectralmodel61}(a), but with different degrees of partial collapsing. (The derivation of Sampler 6 appears in Figure \ref{fig:spectralmodel61} and that of Sampler 10 is omitted to save space.)

The convergence characteristics of $\alpha$, $\beta$, and $\phi$ using Samplers 10 and 11 are compared in Figure \ref{fig:conspectral}; $\gamma$ and $\mu$ converge well for all three samplers. All three MH within PCG samplers outperform the parent Gibbs sampler, since the latter does not converge to the target. Sampler 11 performs much better than Sampler 10 in terms of the mixing and autocorrelations of $\alpha$, $\beta$, and $\phi$. The performance of Sampler 6 is better than Sampler 10, but not as good as Sampler 11. (To save space, the results of the intermediate Sampler 6 are omitted in Figure \ref{fig:conspectral}.) These results show that proper MH within PCG samplers outperform their parent Gibbs sampler in computational efficiency and a higher degree of partial collapsing can improve the convergence even further. 

\subsection{Relating ECME with Newton-type updates to MH within PCG samplers}
\label{sec:enu}

The Expectation-Maximization (EM) algorithm is a frequently used technique for computing maximum likelihood or maximizing a posterior estimate. The Expectation/Conditional Maximization (ECM) algorithm~\citep{meng:rubi:93} extends the EM algorithm by replacing the M-step of each EM iteration with a sequence of CM-steps, each of which maximizes the \emph{constrained} expected complete-data loglikelihood function.~\citet{liu:rubi:94} further generalized ECM with the Expectation/Conditional Maximization Either (ECME) algorithm by replacing some of its CM-steps with steps that maximize the corresponding constrained \emph{actual} likelihood function. ECME can converge substantially faster than either EM or ECM while maintaining the stable monotone convergence and basic simplicity of its parent algorithms. The Gibbs sampler can be viewed as the stochastic counterpart of ECM, see~\citet{vand:meng:10}. PCG extends Gibbs sampling in a manner analogous to ECME's extension of ECM: both PCG and ECME reduce conditioning in a subset of their parameter updates \citep{park:vand:09}. The analogy is not perfect, however. In ECME, for example, the CM-steps maximizing the constrained actual likelihood must be last to guarantee monotone convergence \citep{meng:vand:97}. On the other hand, with PCG, the corresponding partially collapsed steps must be the first to guarantee a proper sampler. 

For ECME, numerical methods, such as Newton-Raphson, may be used to maximize the actual likelihood if no closed-form solution is available. In the context of PCG samplers, these Newton-Raphson steps can often be implemented using MH updates. 

Here we illustrate how this is done by using an ECME algorithm developed for a factor analysis model by \citet{liu:rubi:98}. They derived EM and ECME algorithms and showed that ECME with Newton-type updates converges more quickly than EM. Analogously, it is natural to expect that when fitting this model under a Bayesian framework, a proper MH within PCG sampler will be more efficient than its parent Gibbs sampler. \citet{liu:rubi:98} considered the model,
\begin{eqnarray}
Y_{i}{\sim}{\rm N}_{p}\Big[Z_{i}\beta, \Sigma={\rm Diag}({\sigma}_{1}^{2},\dots,{\sigma}_{p}^{2})\Big],\mbox{ for }i=1,\dots,n,
  \label{eq:factor}
\end{eqnarray}
where $Y_{i}$ is the $(1\times{p})$ vector for observation $i$, $Z_{i}$ is the $(1\times{q})$ vector of the $q$ factors, ${\sigma}_{j}^{2}$  is component $j$ of the diagonal variance-covariance matrix, and $\beta$ is the $(q\times p)$ matrix of factor loadings. We use $\beta_{j}$ to represent column $j$ of $\beta$ and set $Y={\left(Y_1^{\rm T},\dots,Y_n^{\rm T}\right)}^{\rm T}$ and $Z={\left(Z_1^{\rm T},\dots,Z_n^{\rm T}\right)}^{\rm T}$. We use ${\rm N}_{q}(0, I)$ as the prior for $Z_{i}\ (i=1,\dots,n)$ and specify noninformative priors for $\beta$ and $\Sigma$, that is, $p(\sigma_j^2)=\mbox{Inv-Gamma}(0.01,0.01)$ and $p({\beta}_{j})={\rm N}_q\left[0,V={\rm Diag}(100,\dots,100)\right]$ $(j=1,\dots,p)$. \citet{gho:dun:09} discuss this model and its priors in detail.

Sampler 12 (see top panel of Figure~\ref{fig:parentproper}) is a standard Gibbs sampler in which each complete conditional distribution can be sampled directly. To improve its convergence, we construct a proper MH within PCG sampler, Sampler 13, which is also given in Figure~\ref{fig:parentproper}. Because $Z$ is highly correlated with $\sigma_2^2,\dots,\sigma_5^2$, Sampler 13 updates $\sigma_2^2,\dots,\sigma_5^2$ without conditioning on $Z$. Since $\sigma_1^2$ converges  well with the standard Gibbs sampler in the simulation described below, we do not alter its update in Sampler 13. The reduced updates of $\sigma_2^2,\dots,\sigma_5^2$ require MH steps. The derivation of Sampler 13 from its parent Gibbs sampler, i.e., Sampler 12, using the three-phase framework appears in Figure~\ref{fig:factormodel}.

We use a simulation study to illustrate the improved convergence of the MH within PCG sampler over its parent Gibbs sampler. In particular, we set $p=5$, $q=2$, and $n=100$; ${\sigma}_{j}^{2}\ (j=1,\dots,5)$ are generated from Inv-Gamma$(1,0.25)$ and ${\beta}_{hj}\ (h=1,2;j=1,\dots,5)$ from ${\rm N}(0,3^2)$.  We run 20,000 iterations for each sampler with a burnin of 10,000 using the same starting values ($Z_{i}={[1,1]}^{\rm T}$, $\beta_{hj}=1$, and ${\sigma}_{j}^{2}=1$). Figure~\ref{fig:refactor} compares Samplers 12 and 13 in terms of mixing, autocorrelation, and density estimation of ${\sigma}_{2}^{2}$ and ${\sigma}_{3}^{2}$; the first two columns correspond to Sampler 12, and the last two columns correspond to Sampler 13; ${\sigma}_{1}^{2}$ converges well for both samplers, and ${\sigma}_{4}^{2}$ and ${\sigma}_{5}^{2}$ behave similarly as ${\sigma}_{2}^{2}$ and ${\sigma}_{3}^{2}$. The computational advantage of Sampler 13 is evident. More importantly, the MH within PCG sampler delivers a much more trustworth estimate of the marginal posterior distributions as illustrated in the histograms in Figure~\ref{fig:refactor}. 

We repeated the simulation  with $p=50$ and $q=30$ and found that Sampler~13 again outperformed Sampler~12 in a manner similar to what is reported in Figure~\ref{fig:refactor}. When run with $p=50$ and $q=2$, however, both samplers delivered nearly uncorrelated draws.

\section{Discussion}
\label{sec:disc}

Since its introduction in 2008, the PCG sampler has been deployed to improve the convergence properties of numerous Gibbs-type samplers in a variety of applied settings. As with ordinary Gibbs samplers, MH updates are sometimes required within  PCG samplers. Ensuring that the target stationary distribution is maintained in this situation involves subtleties that do not arise in ordinary MH within Gibbs samplers. This has led to the proposal of a number of improper samplers in the literature.  This article elucidates these subtleties, offers a strategy for guaranteeing that the target stationary distribution is maintained, and provides advice as to how best to implement MH within PCG samplers. Some of this advice applies equally to ordinary MH within Gibbs samplers. It is commonly understood, for example, that blocking steps within a Gibbs sampler should improve its convergence. We find, however, that this may not be true if MH is involved.  

{Reducing conditioning in one or more steps of a Gibbs sampler as prescribed by PCG can only improve convergence. If MH is required to implement the reduced steps, however, the overall performance of the algorithm may deteriorate, especially if a poor choice is made for MH jumping rule. Thus, there is a natural trade-off between the computational complexity of MH and the reduced correlation afforded by partial collapsing. Generally speaking, some trial and error may be needed to negotiate this trade-off. In practice we often start with an MH within Gibbs sampler, which already involves MH and can be improved by partial collapsing without any added complexity.} We expect our strategies to extend the application of PCG samplers in practice and to provide researchers with additional tools to improve the convergence of Gibbs-type samplers.

\bigskip

\noindent
{\bf Acknowledgements:} The authors thank Taeyoung Park for helpful comments on a preliminary version of the paper. They also gratefully acknowledge funding for this project partially
provided by the NSF (DMS-12-08791), the Royal Society (Wolfson Merit Award) and the European Commission (Marie-Curie Career Integration Grant).

  
\spacingset{1.4}
\bibliographystyle{natbib}
\bibliography{stats}

\newpage
\spacingset{1.45}
\appendix
\counterwithin{figure}{section}

\bigskip
\begin{center}
{\Large\bf ONLINE SUPPLEMENT: APPENDIX}
\end{center}

\section{Stationary Distribution of Sampler Fragment 6}
\label{sec:figs}

{Figure~\ref{fig:ex3} illustrates how the three-phase framework can be used to verify the stationary distribution of Sampler Fragment 6 of Section~\ref{sec:bnb}, with $\psi_3$ sampled from its complete conditional distribution either before or after Steps~$K$ and $K+1$.}

\vskip 0.4in
\begin{figure}[h]%
\centering
\subfiguretopcaptrue
\subfigure[Parent Gibbs Sampler]{%
\fbox{
\begin{minipage}[c][1.9cm][c]{0.205\linewidth}
\begin{center}
 $ p(\psi_3|\psi_1^\prime,\psi_2^\prime)$\par
 $ p(\psi_2|\psi_1^\prime,\psi_3)$\par
 $ p(\psi_1|\psi_2,\psi_3)$
\end{center}
\end{minipage}
}}%
\hspace{6pt}%
\subfigure[Reduce Conditioning]{%
\fbox{
\begin{minipage}[c][1.9cm][c]{0.205\linewidth}
\begin{center}
 $ p(\psi_3|\psi_1^\prime,\psi_2^\prime)$\par
 $ p(\psi_2^{*}|\psi_1^\prime,\psi_3)$\par
 $\mathcal{M}^\star_{1,2|3}(\psi_1,\psi_2|\psi_1^\prime,\psi_3)$
\end{center}
\end{minipage}
}}
\hspace{0pt}
\subfigure[Permute]{%
\fbox{
\begin{minipage}[c][1.9cm][c]{0.205\linewidth}
 \begin{center}
$ p(\psi_3|\psi_1^\prime,\psi_2^\prime)$\par
$\mathcal{M}^\star_{1,2|3} (\psi_1,\psi_2^{*}|\psi_1^\prime,\psi_3)$\par
$ p(\psi_2|\psi_1,\psi_3)$
\end{center}
\end{minipage}
}}%
\hspace{6pt}%
\subfigure[Trim]{%
\fbox{
\begin{minipage}[c][1.9cm][c]{0.205\linewidth}
 \begin{center}
$ p(\psi_3|\psi_1^\prime,\psi_2^\prime)$\par
$\mathcal{M}_{1|3} (\psi_1|\psi_1^\prime,\psi_3)$\par
$p(\psi_2|\psi_1,\psi_3)$
\end{center}
\end{minipage}
}}\\
\subfigure{%
\fbox{
\begin{minipage}[c][1.9cm][c]{0.205\linewidth}
\begin{center}
 $ p(\psi_2|\psi_1^\prime,\psi_3^\prime)$\par
 $ p(\psi_1|\psi_2,\psi_3^\prime)$\par
 $ p(\psi_3|\psi_1,\psi_2)$
\end{center}
\end{minipage}
}}%
\hspace{6pt}%
\subfigure{%
\fbox{
\begin{minipage}[c][1.9cm][c]{0.205\linewidth}
 \begin{center}
$ p(\psi_2^{*}|\psi_1^\prime,\psi_3^\prime)$\par 
$\mathcal{M}^\star_{1,2|3}(\psi_1,\psi_2|\psi_1^\prime,\psi_3^\prime)$\par
$ p(\psi_3|\psi_1,\psi_2)$
\end{center}
\end{minipage}
}}
\hspace{0pt}
\subfigure{%
\fbox{
\begin{minipage}[c][1.9cm][c]{0.205\linewidth}
 \begin{center}
\vskip -0.1in
 $\mathcal{M}^\star_{1,2|3} (\psi_1,\psi_2^{*}|\psi_1^\prime,\psi_3^\prime)$\par
 $ p(\psi_2|\psi_1,\psi_3^\prime)$\par
 $ p(\psi_3|\psi_1,\psi_2)$
\end{center}
\end{minipage}
}}%
\hspace{6pt}%
\subfigure{%
\fbox{
\begin{minipage}[c][1.9cm][c]{0.205\linewidth}
 \begin{center}
 $\mathcal{M}_{1|3} (\psi_1|\psi_1^\prime,\psi_3^\prime)$\par
 $ p(\psi_2|\psi_1,\psi_3^\prime)$\par
 $ p(\psi_3|\psi_1,\psi_2)$
\end{center}
\end{minipage}
}}%
\caption{Three-phase framework to derive Sampler~Fragment~6 in Section \ref{sec:bnb} from its parent Gibbs sampler. The first row corresponds to updating $\psi_3$ before Steps~$K$ and $K+1$, while the second row updating $\psi_3$ after that.}%
\label{fig:ex3}%
\end{figure}

\section{Details of the Steps in the Gibbs-type Samplers}

This section consists of two parts. The first describes details of sampling steps of the parent Gibbs sampler and proper MH within PCG samplers, i.e., Samplers 6, 10 and 11, for the spectral model~(\ref{eq:sesa}). The second describes the steps of Samplers 12 and 13 which fit the factor analysis model~(\ref{eq:factor}).

\subsection*{B1. Details of the steps in the Gibbs-type samplers based on model~(\ref{eq:sesa})}
\label{ap:specsteps}

Here we assume $X$ is directly observed and we can ignore (\ref{eq:nusesa}) -- (\ref{eq:sa}). With noninformative uniform prior distributions for all of the parameters, the posterior distribution of the parameters $\alpha$, $\beta$, $\gamma$, $\mu$, and $\phi$ under the spectral model~(\ref{eq:sesa}) is
$$p(\alpha,\beta,\gamma,\mu,\phi|X)\propto\displaystyle{\prod_{i=1}^n {\left[\alpha({E_{i}}^{-\beta}+{\gamma}I\{i=\mu\})e^{-\phi/E_{i}}\right]}^{X_{i}}{\rm exp}\left\{ -\alpha\sum_{i=1}^n({E_{i}}^{-\beta}+{\gamma}I\{i=\mu\})e^{-\phi/E_{i}}\right\}}.\eqno{({\rm B}1.1)}$$

\noindent The joint posterior distribution of the parameters and augmented data $X_L$ is
$$
\begin{array}{ll}
p(\alpha,\beta,\gamma,\mu,\phi,X_{L}|X)\propto & \displaystyle{\alpha^{\sum_{i=1}^n X_{i}}e^{-\phi\sum_{i=1}^n (X_{i}/E_{i})}\prod_{i=1}^n E_{i}^{-\beta (X_i-X_{iL})} \gamma^{\sum_{i=1}^n X_{iL}}\times}\\ &\displaystyle{\prod_{i=1}^n{\left\{I(i=\mu)\right\}}^{X_{iL}}{\rm exp}\left\{-\alpha\sum_{i=1}^n({E_{i}}^{-\beta}+{\gamma}I\{i=\mu\})e^{-\phi/E_{i}}\right\}}.
\end{array}\eqno{({\rm B}1.2)}$$
Thus the steps of the parent MH within Gibbs sampler in Figure \ref{fig:spectralmodel61}(a) or \ref{fig:spectralmodel111}(a) are 
\begin{steps}
\itemsep=0in
\step Sample $X_{iL}$ from ${\rm Binomial}\displaystyle{\left\{X_i,\frac{{\gamma}I\{i=\mu\}}{{E_{i}}^{-\beta}+{\gamma}I\{i=\mu\}}\right\}}$, for $i=1,\dots,n$,
\step Sample $\alpha$ from ${\rm Gamma}\displaystyle{\left\{\sum_{i=1}^n X_{i}+1,\sum_{i=1}^n({E_{i}}^{-\beta}+{\gamma}I\{i=\mu\})e^{-\phi/E_{i}}\right\}}$,
\step Use MH to sample $\beta$ from $p(\beta|X,X_{L},\alpha,\gamma,\mu,\phi)\propto p(\alpha,\beta,\gamma,\mu,\phi,X_{L}|X)$,
\step  Sample $\gamma$ from ${\rm Gamma}\displaystyle{\left\{\sum_{i=1}^n X_{iL}+1,\alpha\sum_{i=1}^n I\{i=\mu\}e^{-\phi/E_{i}}\right\}}$,
\step Use MH to sample $\mu$ from $p(\mu|X,X_L,\alpha,\beta,\gamma,\phi)\propto p(\alpha,\beta,\gamma,\mu,\phi,X_L|X)$,
\step Use MH to sample $\phi$ from $p(\phi|X,X_{L},\alpha,\beta,\gamma,\mu)\propto p(\alpha,\beta,\gamma,\mu,\phi,X_{L}|X)$.
\end{steps}
\noindent The steps of Sampler 10 are
\begin{steps}
\itemsep=0in
\step Use MH to sample $\mu$ from $p(\mu|X,\alpha,\beta,\gamma,\phi)\propto p(\alpha,\beta,\gamma,\mu,\phi|X)$,
\step Sample $X_{iL}$ from ${\rm Binomial}\displaystyle{\left\{X_i,\frac{{\gamma}I\{i=\mu\}}{{E_{i}}^{-\beta}+{\gamma}I\{i=\mu\}}\right\}}$, for $i=1,\dots,n$,
\step Sample $\alpha$ from ${\rm Gamma}\displaystyle{\left\{\sum_{i=1}^n X_{i}+1,\sum_{i=1}^n({E_{i}}^{-\beta}+{\gamma}I\{i=\mu\})e^{-\phi/E_{i}}\right\}}$,
\step Use MH to sample $\beta$ from $p(\beta|X,X_{L},\alpha,\gamma,\mu,\phi)\propto p(\alpha,\beta,\gamma,\mu,\phi,X_{L}|X)$,
\step  Sample $\gamma$ from ${\rm Gamma}\displaystyle{\left\{\sum_{i=1}^n X_{iL}+1,\alpha\sum_{i=1}^n I\{i=\mu\}e^{-\phi/E_{i}}\right\}}$,
\step Use MH to sample $\phi$ from $p(\phi|X,X_{L},\alpha,\beta,\gamma,\mu)\propto p(\alpha,\beta,\gamma,\mu,\phi,X_{L}|X)$.
\end{steps}

\noindent Integrating $({\rm B}1.1)$ over $\alpha$, we have,
$$\begin{array}{ll}p(\beta,\gamma,\mu,\phi|X)\propto&\displaystyle{\prod_{i=1}^n {\left[({E_{i}}^{-\beta}+{\gamma}I\{i=\mu\})e^{-\phi/E_{i}}\right]}^{X_{i}}\times}\\&\displaystyle{{\left[\sum_{i=1}^n({E_{i}}^{-\beta}+{\gamma}I\{i=\mu\})e^{-\phi/E_{i}}\right]}^{-(\sum_{i=1}^n X_{i}+1)}}.\end{array}\eqno{({\rm B}1.3)}$$

\noindent Hence, the steps of Sampler 6 are
\begin{steps}
\itemsep=0in
\step Use MH to sample $\mu$ from $p(\mu|X,\beta,\gamma,\phi)\propto p(\beta,\gamma,\mu,\phi|X)$,
\step Use MH to sample $\phi$ from $p(\phi|X,\beta,\gamma,\mu)\propto p(\beta,\gamma,\mu,\phi|X)$,
\step Use MH to sample $\beta$ from $p(\beta|X,\gamma,\mu,\phi)\propto p(\beta,\gamma,\mu,\phi|X)$,
\step Sample $\alpha$ from ${\rm Gamma}\displaystyle{\left\{\sum_{i=1}^n X_{i}+1,\sum_{i=1}^n({E_{i}}^{-\beta}+{\gamma}I\{i=\mu\})e^{-\phi/E_{i}}\right\}}$,
\step Sample $X_{iL}$ from ${\rm Bin}\displaystyle{\left\{X_i,\frac{{\gamma}I\{i=\mu\}}{{E_{i}}^{-\beta}+{\gamma}I\{i=\mu\}}\right\}}$, for $i=1,\dots,n$,
\step  Sample $\gamma$ from ${\rm Gamma}\displaystyle{\left\{\sum_{i=1}^n X_{iL}+1,\alpha\sum_{i=1}^n I\{i=\mu\}e^{-\phi/E_{i}}\right\}}$.
\end{steps}

\noindent The steps of Sampler 11 are almost the same as Sampler 6, except Steps 2 and 3 are combined into one step. That is, we use MH to sample $(\beta,\phi)$ from $p(\beta,\phi|X,\gamma,\mu)\propto p(\beta,\gamma,\mu,\phi|X)$. 
 
We use a uniform distribution on $\{1,\dots,n\}$ as the  jumping rule when updating $\mu$. When updating either $\beta$ or $\phi$ via MH, we use a normal distribution centered at the current draw of the parameter for the jumping rule; the variance of the jumping rule is 
adjusted to obtain an acceptance rate of around $40\%$. Analogously, when sampling $\beta$ and $\phi$ jointly via MH, the jumping rule is a bivariate normal distribution centered at the current draw with variance-covariance matrix adjusted to obtain an acceptance rate of around $20\%$. 

\subsection*{B2. Details of the steps in the Gibbs-type samplers based on model~(\ref{eq:factor})}
\label{ap:factsteps}

With priors $p(\sigma_j^2)=\mbox{Inv-Gamma}(a,b)$ and $p(\beta_j)={\rm N}_2\left(0,V\right)$ $(j=1,\dots,5)$, the posterior distribution of the parameters $Z$, $\beta$, and $\Sigma$ under the factor analysis model~(\ref{eq:factor}) is
$$\begin{array}{lll}p(Z,\beta,\Sigma|Y)&\propto&\displaystyle{{\left|\Sigma\right|}^{-n/2}{\left(\prod_{j=1}^5 \sigma_j^{-2(a+1)}\right)}{\rm exp}\left\{ -\frac{1}{2}\sum_{i=1}^n\left[(Y_i-Z_i\beta)\Sigma^{-1}{(Y_i-Z_i\beta)}^{\rm T} +Z_i {Z_i}^{\rm T}\right]\right\}}\\&&{\rm exp}\displaystyle{\left\{-\frac{1}{2}\sum_{j=1}^5 \beta_j^{\rm T} V^{-1} \beta_j-b\sum_{j=1}^5 \sigma_j^{\rm -2}\right\}}.\end{array}\eqno{({\rm B}2.1)}$$
Thus the steps of Sampler 12 are
\begin{description}
\itemsep=0in
\item[Step $1$:] Sample $Z_{i}$ from $\displaystyle{{\rm N}_2\left[{(I_2+\beta\Sigma^{-1}\beta^{\rm T})}^{-1}\beta\Sigma^{-1}Y_i^{\rm T},{(I_2+\beta\Sigma^{-1}\beta^{\rm T})}^{-1}\right]}$, for $i=1,\dots,100$,
\item[Step $2$:]  Sample $\sigma_{j}^{2}$ from Inv-Gamma$\displaystyle{\left\{a+\frac{n}{2},b+\frac{1}{2}\sum_{i=1}^n {(Y_{ij}-\beta_j^{\rm T} Z_i^{\rm T})}^2\right\}}$, for $j=1,\dots,5$,
\item[Step $3$:] Sample $\beta_{j}$ from $\displaystyle{{\rm N}_2\left[({V^{-1}+Z^{\rm T}Z/\sigma_j^2)}^{-1}Z^{\rm T} Y_{.j}/\sigma_j^2,{(V^{-1}+Z^{\rm T}Z/\sigma_j^2)}^{-1}\right]}$, for $j=1,\dots,5$,
\end{description}
where $Y_{.j}$ represents the $j$th column of $Y$. Integrating $({\rm B}2.1)$ over $Z$, we have,
$$\begin{array}{lll}p(\beta,\Sigma|Y)&\propto&\displaystyle{{\left|I_2+\beta\Sigma^{-1}\beta^{\rm T}\right|}^{-n/2}{\left|\Sigma\right|}^{-n/2}{\rm exp}\left\{ -\frac{1}{2}\sum_{i=1}^n\left[Y_i(\Sigma^{-1}-\Sigma^{-1}\beta^{\rm T}{(I_2+\beta\Sigma^{-1}\beta^{\rm T})}^{-1}\beta\Sigma^{-1})Y_i^{\rm T} \right]\right\}}\\&&\displaystyle{{\left(\prod_{j=1}^5 \sigma_j^{-2(a+1)}\right)}}{\rm exp}\displaystyle{\left\{-\frac{1}{2}\sum_{j=1}^5 \beta_j^{\rm T} V^{-1} \beta_j-b\sum_{j=1}^5 \sigma_j^{\rm -2}\right\}}.\end{array}\eqno{({\rm B}2.2)}$$

\noindent Hence, the steps of Sampler 13 are
\begin{description}
\itemsep=0in
\item[Step $1$:] Sample $\sigma_{1}^{2}$ from Inv-Gamma$\displaystyle{\left\{a+\frac{n}{2},b+\frac{1}{2}\sum_{i=1}^n {(Y_{i1}-\beta_1^{\rm T} Z_i^{\rm T})}^2\right\}}$,
\item[Step $j$:] Use MH to sample $\sigma_{j}^{2}$ from $p(\sigma_{j}^2|Y,\beta,\sigma_{1}^{2},\dots,\sigma_{j-1}^{2},\sigma_{j+1}^{2},\dots,\sigma_{5}^{2})\propto p(\beta,\Sigma|Y)$, for $j=2,\dots,5$, 
\item[Step $6$:] Sample $Z_{i}$ from $\displaystyle{{\rm N}_2\left[{(I_2+\beta\Sigma^{-1}\beta^{\rm T})}^{-1}\beta\Sigma^{-1}Y_i^{\rm T},{(I_2+\beta\Sigma^{-1}\beta^{\rm T})}^{-1}\right]}$, for $i=1,\dots,100$,
\item[Step $7$:] Sample $\beta_{j}$ from $\displaystyle{{\rm N}_2\left[({V^{-1}+Z^{\rm T}Z/\sigma_j^2)}^{-1}Z^{\rm T} Y_{.j}/\sigma_j^2,{(V^{-1}+Z^{\rm T}Z/\sigma_j^2)}^{-1}\right]}$, for $j=1,\dots,5$.

\end{description}

When updating $\sigma_j^2$ $(j=2,\dots,5)$ via MH, we use a log-normal distribution centered at the log of the current value of the parameter for the jumping rule; the variance  is adjusted to obtain an acceptance rate of around $40\%$.

\end{document}